\documentclass[fleqn,usenatbib]{mnras}

\usepackage[T1]{fontenc}

\DeclareRobustCommand{\VAN}[3]{#2}
\let\VANthebibliography\thebibliography
\def\thebibliography{\DeclareRobustCommand{\VAN}[3]{##3}\VANthebibliography}

\usepackage{graphicx}	
\usepackage{amsmath}	
\usepackage{amssymb}	
\usepackage{graphics}
\usepackage{siunitx} 

\usepackage{newtxtext,newtxmath}

\title[JWST's first glimpse of the EoR]{Seeing sharper and deeper: JWST's first glimpse of the photometric and spectroscopic properties of galaxies in the epoch of reionisation}

\author[J. A. A. Trussler et al.]{James A.\@ A.\@ Trussler,$^{1}$\thanks{E-mail: james.trussler@manchester.ac.uk}
Nathan J.\@ Adams,$^{1}$,
Christopher J.\@ Conselice,$^{1}$
Leonardo Ferreira,$^{2}$
\newauthor
Duncan Austin,$^{1}$
Rachana Bhatawdekar,$^{3}$
Joseph Caruana,$^{4,5}$
Brenda L.\@ Frye,$^{6}$
Tom Harvey,$^{1}$ 
\newauthor
Christopher C.\@ Lovell,$^{7}$ 
Massimo Pascale,$^{8}$
William J.\@ Roper,$^{9}$ 
Aprajita Verma,$^{10}$
Aswin P.\@ Vijayan$^{11,12}$
\newauthor
and Stephen M.\@ Wilkins$^{9,5}$
\\
$^{1}$Jodrell Bank Centre for Astrophysics, University of Manchester, Oxford Road, Manchester M13 9PL, UK\\
$^{2}$Department of Physics and Astronomy, University of Nottingham, NG7 2RD, UK \\
$^{3}$European Space Agency, ESA/ESTEC, Keplerlaan 1, 2201 AZ Noordwijk, NL, The Netherlands\\
$^4$Department of Physics, University of Malta, MSD 2080, Malta \\
$^{5}$Institute of Space Sciences and Astronomy, University of Malta, Msida MSD 2080, Malta \\
$^{6}$University of Arizona, Department of Astronomy/Steward Observatory, 933 N Cherry Ave, Tucson, AZ 8572, USA\\
$^{7}$Centre for Astrophysics Research,  School of Physics, Engineering \& Computer Science, University of Hertfordshire, Hatfield AL10 9AB, UK\\
$^{8}$Department of Astronomy, University of California, 501 Campbell Hall \#3411, Berkeley, CA 94720, USA\\
$^{9}$Astronomy Centre, University of Sussex, Falmer, Brighton BN1 9QH, UK\\
$^{10}$Oxford Astrophysics, University of Oxford, Keble Road, Oxford OX1 3RH, UK \\
$^{11}$Cosmic Dawn Center (DAWN)\\ 
$^{12}$DTU-Space, Technical University of Denmark, Elektrovej 327, DK-2800 Kgs. Lyngby, Denmark \\
}

\date{Accepted XXX. Received YYY; in original form ZZZ}

\pubyear{2023}

\begin{document}
\label{firstpage}
\pagerange{\pageref{firstpage}--\pageref{lastpage}}
\maketitle

\begin{abstract}
We analyse the photometric and spectroscopic properties of four galaxies in the epoch of reionisation (EoR) within the SMACS J0723.3-7327 JWST Early Release Observations field. Given the known spectroscopic redshifts of these sources, we investigated the accuracy with which photometric redshifts can be derived using NIRCam photometry alone, finding that F115W imaging is essential to distinguish between $z\sim8$ galaxies with high equivalent width (EW) [\ion{O}{III}] $\lambda 5007$ emission and $z\sim10$ Balmer break galaxies. We find that all four sources exhibit strong ($\geq$ 0.6~mag) F356W$-$F444W colours, which sit at the extreme end of theoretical predictions from numerical simulations. We find that these galaxies deviate (by $\approx$ 0.5~dex) from the local correlation between [\ion{O}{III}] $\lambda 5007$/H$\beta$ and [\ion{Ne}{III}] $\lambda 3869$/[\ion{O}{II}], which is consistent with the predictions from simulations of high-redshift galaxies having elevated line excitation ratios. We measure the [\ion{O}{III}] $\lambda 5007$ rest-frame equivalent widths both directly from the spectroscopy, and indirectly as inferred from the strong F356W$-$F444W colours, finding large [\ion{O}{III}] $\lambda 5007$ EWs of 225--1740~\AA. The [\ion{O}{III}] $\lambda 5007$ and H$\beta$ EWs are consistent with those seen in extreme, intensely star-forming dwarf galaxies in the local Universe. Our structural analysis indicates that these galaxies are resolved, exhibiting irregular shapes with bright clumps. In line with the predictions from the FLARES hydrodynamic simulations, such intense star formation and extreme nebular conditions are likely the norm, rather than the exception, in the EoR. 

\end{abstract}




\section{Introduction}

A new eye on our Universe has opened. And it is a golden eye, for the \emph{James Webb Space Telescope} (\emph{JWST}) has commenced its mission, to survey the depths of the night sky on our behalf. In doing so, it will greatly build on the legacy of its predecessor telescopes, capturing the complexity and grandeur of the Universe with a sharpness of vision and depth that is unlike anything we have seen before. Indeed, \emph{JWST} will build on the legacy of the \emph{Hubble Space Telescope}, which, over the years, after generation after generation of new instrumentation, from NICMOS to ACS to WFC3, has steadily pushed the frontiers of the study of the formation and evolution of galaxies to increasingly higher redshifts \citep[see e.g.\@][]{Ferguson2000, Bromm2011, Dunlop2013, Duncan2014, Finkelstein2016}. But \emph{JWST} will see even sharper and deeper still, bringing us ever closer to witnessing the assembly of the very first galaxies at the dawn of time. However, it is also the legacy of the \emph{Spitzer Space Telescope} that \emph{JWST} will be building upon. Indeed, with its infrared vision, \emph{Spitzer} has enabled us to see galaxies within the epoch of reionisation in the rest-frame optical, the very light that formed the foundation of astronomy over the ages, but at a lower resolution.

It is precisely this rest-frame optical light that yields such valuable insights into galaxy formation and evolution. Unlike the rest-frame UV light that \emph{Hubble} probes in the epoch of reionisation \citep[e.g.,][]{Bouwens2015, Finkelstein2015, Bhatawdekar2019} which is primarily an indicator for young stellar populations, the rest-frame optical probed by \emph{Spitzer} is also sensitive to older, more evolved stellar populations \citep{Maraston2011, Conroy2013a}. Indeed, as seen from spectroscopy, the rest-frame optical contains a wealth of information on the conditions within galaxies. For example, measurements of nebular emission lines can give insights on the star formation rates within galaxies \citep{Kennicutt1998a, Kennicutt2012}, the primary excitation mechanisms responsible for energising the ISM \citep{Baldwin1981, Kewley2001, Steidel2014, Shapley2015a, Belfiore2016, Curti2022a}, the gas density \citep{Mingozzi2019, Fluetsch2021} and amount of dust \citep{Calzetti2000, Charlot2000}, as well as the chemistry \citep{Tremonti2004, Mannucci2010, Curti2017, Maiolino2019} and ionisation state of the gas \citep{Kumari2019, Mingozzi2020}. Furthermore, through a measurement of the stellar continuum, it is possible to put constraints on the stellar populations within galaxies, such as the chemistry of the stars \citep{Peng2015, Kriek2016, Trussler2020, Trussler2021, Carnall2022} and the star formation histories and stellar ages of galaxies \citep{Gallazzi2005, Chauke2018, Carnall2019}. 

Photometry from \emph{Spitzer}/IRAC imaging delivered valuable insights into the nebular conditions within galaxies. Indeed, galaxies in the EoR were found to have prominent flux excesses in one of the two IRAC 3.6 or 4.5 \textmu m bands, thought to be driven by high equivalent width [\ion{O}{III}] $\lambda 5007$ and H$\beta$ emission, boosting up the fluxes well beyond the continuum level \citep{Schaerer2009, Laporte2014, Smit2014, Smit2015, Endsley2021}. The intense star formation inferred for these EoR galaxies was therefore akin to that seen in the extreme, vigorously star-forming dwarf galaxies in the local Universe \citep[e.g.,][]{Searle1972, Izotov2011, James2016}. Such intense star formation in EoR galaxies does pose a problem, for it becomes challenging to accurately recover their properties, such as stellar masses from the photometry \citep{Smit2014}. Indeed, one innovative method to separate the nebular emission and the starlight is to intentionally study galaxies at sufficiently high redshifts such that the bright nebular lines become redshifted out of the \emph{Spitzer} bands \citep{Laporte2021}. In doing so, tighter constraints on the star formation histories and ages of EoR galaxies are possible, with \citet{Laporte2021} finding that the IRAC excess seen in some $z > 9$ galaxies may be attributable to a Balmer break, i.e.\@ a recent lack of star formation. Hence in order to better understand the star formation (and potential quenching) in EoR galaxies, a clean separation between starlight and nebular emission is essential.

\emph{JWST}/NIRSpec, with its unique spectroscopic capabilities in the near-infrared, now enables us to probe the rest-frame optical of EoR galaxies through spectroscopy for the first time \citep[see e.g.\@][]{Brinchmann2022, Carnall2023, Curti2023, Schaerer2022, Tacchella2022, Trump2022}. It is thus the ideal facility with which to characterise the physical properties of these galaxies, and to separate the nebular emission from the starlight \citep{Gardner2006, Bromm2011, Finkelstein2016}. Through it, we will be able to directly and immediately establish whether the prominent IRAC excess is driven by star formation \cite[or a lack thereof,][]{Smit2014, Smit2015, Laporte2021}. Furthermore, in knowing both the photometry and the spectroscopic redshifts of EoR galaxies, we will be able to determine how effectively galaxies in the epoch of reionisation actually can be identified from photometric redshifts derived from fits to their SEDs and through colour--colour selections.

In this paper we aim to draw on the synergy between photometry and spectroscopy, by jointly analysing both NIRCam imaging and NIRSpec spectroscopic data for the four EoR galaxies that were observed in the \emph{JWST} Early Release Observations (ERO) of the SMACS J0723.3-732 (SMACS 0723 hereafter) cluster field. Our analysis will thus focus on both the photometric and spectroscopic properties of these galaxies. Leveraging on the unique spectroscopic capabilities of NIRSpec, which enables us to directly probe the rest-frame optical properties of galaxies deep within the epoch of reionisation, we apply emission line diagnostics to constrain the ISM conditions and star formation activity within these systems. We compare the characteristics of the EoR sources to extreme, intensely star-forming dwarf galaxies in the local Universe to determine to what extent the galaxy population has evolved from the epoch of reionisation to the present day. In particular, we establish to what degree the IRAC excess (now indicated by red NIRCam F356W$-$F444W colours) is attributable to high equivalent width [\ion{O}{III}] $\lambda 5007$ emission, and how consistent the equivalent widths traditionally inferred from photometry are with the now newly available direct measurements from spectroscopy. Furthermore, with the spectroscopic redshifts of these sources being known, we can now undertake a preliminary investigation of the actual prospects of truly identifying such sources from SED fitting and colour--colour selection applied solely to NIRCam photometry. Additionally, we intend to demonstrate the complementary power of NIRCam imaging and NIRSpec spectroscopy, which together will help to distinguish between nebular emission driven by star formation and that which arises from AGN activity.

This paper is structured as follows. In Section~\ref{sec:data}, we discuss the NIRCam imaging and NIRSpec spectroscopy that we will use in our analysis. In Section~\ref{sec:photometry}, we analyse the photometric properties of the four EoR galaxies, focussing on their morphologies, and the prospects for reliably identifying such galaxies from SED fitting and colour selection applied to NIRCam photometry. In Section~\ref{sec:spectroscopy}, we investigate the spectroscopic properties of these galaxies, focussing on their emission line ratios and equivalent widths, to place constraints on the nebular conditions within these galaxies, as well as their star formation activity. Finally, in Section~\ref{sec:conclusions} we summarise our findings and conclude. 

\section{Data} \label{sec:data}

\begin{figure*}
\centering
\includegraphics[width=.7\linewidth]{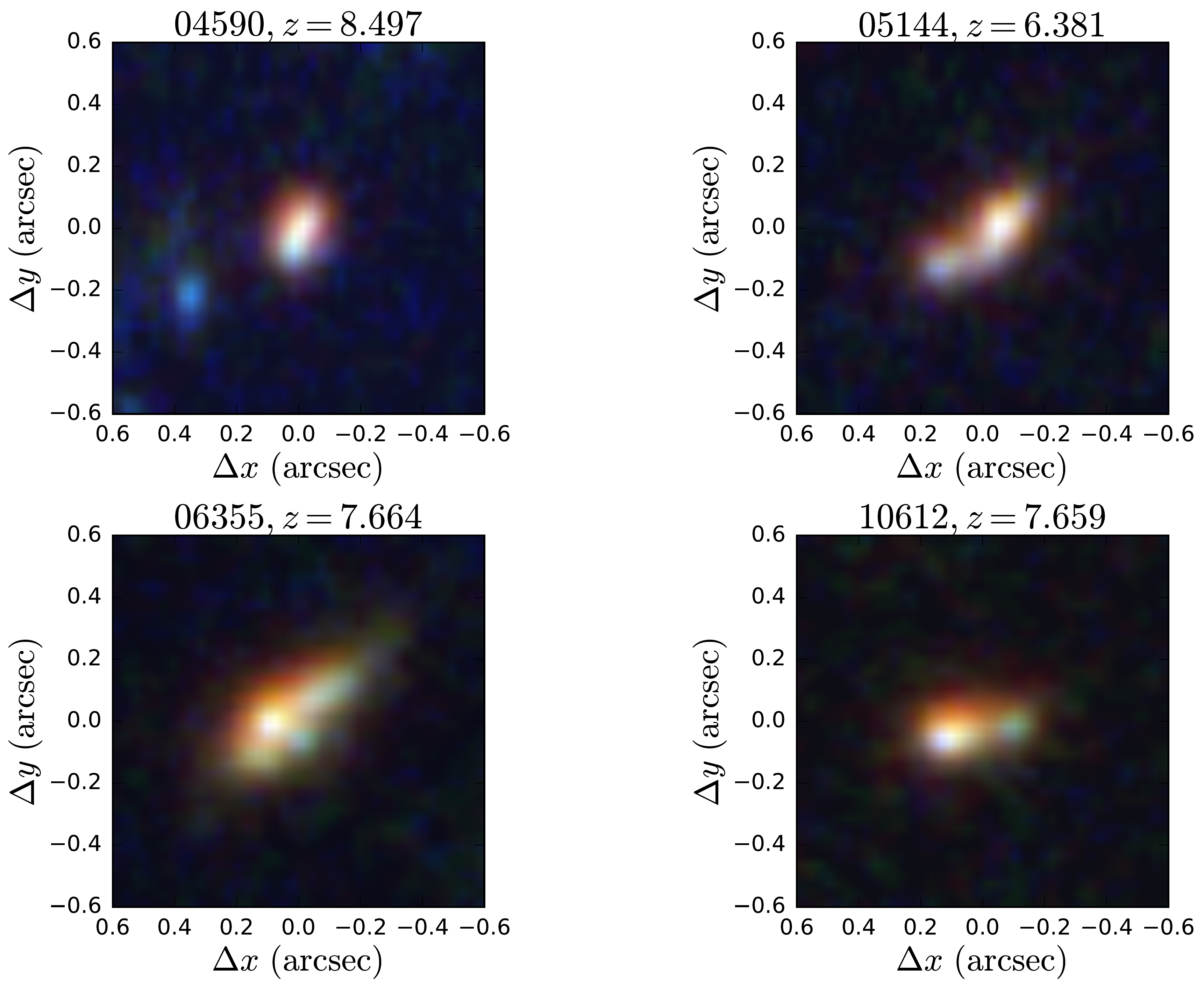}
\caption{RGB cutouts \citep[B: F090W, F150W; G: F200W, F277W; R: F356W, F444W, generated using Trilogy,][]{Coe2012} for the four epoch of reionisation galaxies in the SMACS 0723 field that were targeted for follow-up spectroscopy with NIRSpec. These galaxies are resolved with NIRCam, exhibiting complex, irregular shapes (perhaps in part due to distortions by gravitational lensing) with bright clumps. Thus the standard flux calibration procedures for reducing NIRSpec spectra (which assume either point-source or uniform morphologies) will likely not be optimal, even for these high-$z$ objects. Note that we have not PSF-matched the images in the various filters when generating the RGB cutouts, so as to highlight the irregular, clumpy nature of these objects, which is seen more clearly in the shorter wavelength NIRCam bands.}
\label{fig:eor_cutouts}
\end{figure*}

\begin{table*}
\caption{The rest-frame UV and rest-frame optical morphological parameters of the four EoR galaxies studied in this work. We provide both the effective radius $R_\mathrm{e}$ and asymmetry parameters $A$, as well as the Sérsic index $n$. The spectroscopic redshift $z$, stellar mass $M_*$ and star formation rate $\mathrm{SFR}$ (both derived following the procedure in Section~\ref{subsec:photometry}, having fixed the redshift at the spectroscopic redshift, using FLUX\_AUTO fluxes from SExtractor, and corrected for magnification) and magnification factor $\mu$ \citep[using the \emph{JWST}-based magnification maps from][]{Pascale2022} are also provided.}
\centering
\begin{tabular}{cccccccccc}
\hline
ID & $z$  &  $\rm UV \ R_{e} \ [kpc]$ & $\rm Optical \ R_{e} \ [kpc]$ & $\rm n({\rm Optical)}$ & $\rm A(UV)  $ & A($\rm Optical)$ & $\log (M_*/\mathrm{M}_\odot)$ & $\mathrm{SFR} \ \mathrm{[M_\odot /yr]}$& $\mu $ \\ \hline
04590 & 8.497   & $0.69 \pm  0.08 $ & $0.48 \pm 0.01$ & $1.72 \pm 0.12$ & $0.12 \pm 0.10 $ & $0.38 \pm 0.06$ & $8.16 \pm 0.65$ & $1.30 \pm 0.96$ & $5.8$ \\
05144 & 6.381   & $1.03 \pm 0.02$ & $0.87 \pm 0.02$ & $0.67 \pm 0.04$ & $0.27 \pm 0.05$ & $0.24 \pm 0.05$ & $7.56 \pm 0.15$ & $0.37 \pm 0.17$ & $3.2$ \\
06355 & 7.664   & $0.97 \pm 0.11$ & $0.88 \pm 0.02$ & $0.66 \pm 0.06$ & $0.23 \pm 0.20$ & $0.23 \pm 0.05$ & $8.65 \pm 0.09$ & $4.47 \pm 1.04$ & $1.6$ \\
10612 & 7.659  & $0.65 \pm 0.10$ & $0.73 \pm 0.03$ & $1.10 \pm 0.05$ & $0.23 \pm 0.03$ & $0.35 \pm 0.06 $ & $7.90 \pm 0.14$ & $0.79 \pm 0.32$ & $1.7$\\
       \hline
       \label{tab:structural_properties}
\end{tabular}
\end{table*}

We make use of both the NIRCam \citep{Rieke2005} imaging and NIRSpec \citep{Ferruit2022, Jakobsen2022} spectroscopic data taken during the \emph{JWST} Early Release Observations \citep[PI: Pontopiddan, Program ID: 2736,][]{Pontoppidan2022} of the the RELICS cluster J0723.3-732 \citep[SMACS 0723,][]{Coe2019, Golubchik2022, Pascale2022}. Our data reduction procedure is as follows.

In terms of the NIRCam data, the imaging in the F090W, F150W and F200W short wavelength bands, together with the F277W, F356W and F444W long wavelength bands, were obtained on June 7, 2022. The total integration time for this field was 12.5~hr. The NIRCam data was processed following the reduction steps outlined in more detail in \citet{Ferreira2022}, \citet{Adams2023} and \citet{Austin2023}. Briefly, the uncalibrated lower-level \emph{JWST} data products for this field were reprocessed following a modified version of the \emph{JWST} official pipeline (pipeline version 1.8.2 and Calibration Reference Data
System (CRDS) context jwst\_0995.pmap), as the initial release of the higher-level products were found to contain WCS alignment issues as well as sub-optimal background subtraction. Sources were identified and extracted using {\footnotesize SExtractor} \citep{Bertin1996}, with our catalog being generated by applying forced photometry on the images, using the F444W band as the selection band. In our SED fitting and colour analysis in this paper, we make use of forced aperture photometry that was calculated within $0.32''$ diameter circular apertures. These aperture magnitudes are corrected using an aperture correction derived from simulated WebbPSF point spread functions for each band used \citep{Adams2023}. For our determination of total stellar masses and star formation rates (computed using {\tt BAGPIPES}, tabulated in Table~\ref{tab:structural_properties} and discussed throughout this work), we instead use Kron elliptical apertures (i.e.\@ the FLUX\_AUTO fluxes in {\footnotesize SExtractor}), which are based off of the F444W detection band. These extraction apertures, together with the cutouts of our sources in the six available NIRCam bands, are shown in Appendix~\ref{app:photometry}.

The NIRSpec data we use were taken on June 30, 2022, and were split between two observations (007 and 008). In each observation, 2.5~h integrations in both the G235M and G395M NIRSpec gratings were obtained. We make use of the default reductions of the NIRSpec spectra that are publicly available on MAST (pipeline version 1.8.2 and CRDS context jwst\_1041.pmap). Given the presence of ``snowballs'' in the NIRSpec detector images, which are likely caused by cosmic ray events, and are currently not well-removed by the default configuration of the \emph{JWST}/NIRSpec reduction pipeline, there are notable spikes and troughs in the default reductions of the NIRSpec spectra used in this analysis. The spectral regions affected by cosmic rays (inferred by eye) have therefore been masked and are excluded from our analysis. 
 
 We focus our analysis on the spectra obtained using the NIRSpec/G395M grating, which covers the spectral range $2.9 < \lambda$ (\textmu m) $ < 5.3$ at a spectral resolution $R\sim1000$. We do not incorporate the spectra taken by the NIRSpec/G235M grating, which covers the spectral range $1.7 < \lambda$ (\textmu m) $ < 3.2$ at $R\sim1000$. We adopt this approach for two reasons. Firstly, because the G395M grating is $\approx 50\%$ more sensitive than the G235M grating. Secondly, because the bright rest-frame optical lines, which are sensitive tracers to the conditions within the ISM, fall into the spectral range of the G395M grating at these redshifts. 

\section{Photometric properties of the EoR galaxies} \label{sec:photometry}

\subsection{The structures of EoR galaxies, as seen through JWST}

We show the RGB cutouts \citep[B: F090W, F150W; G: F200W, F277W; R: F356W, F444W, generated using Trilogy,][]{Coe2012} for the four EoR galaxies that were targeted for follow-up NIRSpec spectroscopy in Fig.\@~\ref{fig:eor_cutouts}. We note that, as these four galaxies reside in the SMACS 0723 cluster field, they are subject to magnification (with magnification factors shown in Table~\ref{tab:structural_properties}), with their spatial profiles likely having been distorted to some degree by gravitational lensing \citep{Pascale2022}. We do not correct the morphologies for gravitational lensing in this work. Thus the morphological parameters we derive only represent the perceived properties of these galaxies, rather than their true, inherent physical sizes and shapes. 

\begin{figure}
\centering
\includegraphics[width=0.45\textwidth]{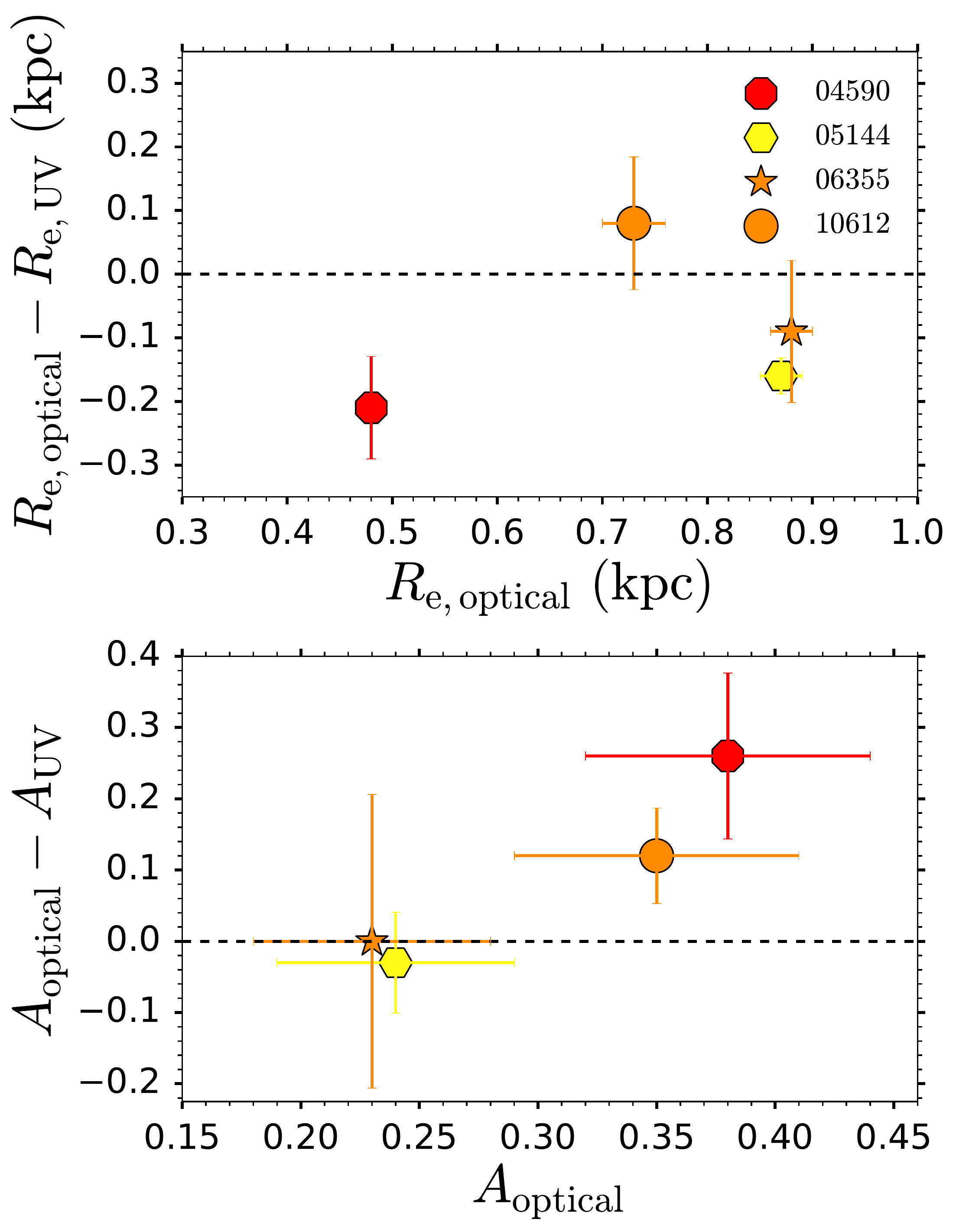}
\caption{Size and asymmetry offsets between optical and UV observations. Top: The difference in effective radius ($R_e$) between the rest-frame optical (probed by F444W) and the rest-frame UV (probed by F200W for the three $z > 7$ galaxies, F150W for the $z=6.381$ galaxy 05144) for the 4 sources presented. The UV sizes are overall larger than the optical with the exception of ID~10612 that shows no clear offset. Bottom: Asymmetry ($A$) offsets between the rest-frame optical and UV. Two sources (ID~04590, 10612) show more irregular and clumpy regions in the optical images, while ID~05144 and ID~06355 show similar asymmetries between both wavelengths. Note that while the images (and hence the morphological parameters measured) have not been corrected for gravitational lensing, they have all been PSF-matched to the F444W filter.}

\label{fig:sizes_asymmetry}
\end{figure}

We used the \textsc{MORFOMETRYKA} code \citep{Ferrari2015, Ferreira2022} to perform a structural analysis of these sources by fitting a single component Sérsic profile to their 2D light distributions and measuring quantitative morphology parameters such as the asymmetry ($A$). A detailed description of the algorithm and measurements can be found in \cite{Ferrari2015}. The morphological parameters that were derived are provided in Table~\ref{tab:structural_properties}. As can be seen from the cutouts, the four EoR galaxies are clearly resolved and have distinct, irregular shapes with bright clumps. These clumpy structures are revealed clearly in Sérsic residuals after model subtraction, characterized by concentrated nodes not following a regular smooth light profile. We show the difference between the rest-frame UV (probed by F200W for the three $z>7$ galaxies, F150W for the $z=6.381$ galaxy 05144) and the rest-frame optical sizes (probed by F444W) and asymmetries in Fig.\@~\ref{fig:sizes_asymmetry}, where the images have all been PSF-matched to the F444W filter. Three out of the four sources (ID~04590, 05144, 10612) display larger UV sizes, with the exception of ID~10612, which has a UV size consistent with the optical measurement within the error bars. The asymmetries between UV and optical are consistent for two sources (ID~05144, 06355), while ID~04590 and ID~10612 show more asymmetric optical light, with $|\Delta A| \sim 0.26$ and $|\Delta A| \sim 0.12$ respectively. 


As we will comment further in Section~\ref{sec:spectroscopy}, these galaxies have neither point-source nor uniform morphologies \citep[in agreement with the clumpy structures seen with \emph{HST} imaging, e.g.\@][]{Jiang2013, Bowler2017a} implying that the standard flux calibration procedures for reducing NIRSpec spectra will not be optimal, even for these high-$z$ objects. 

We note, as has already been pointed out in \citet{Carnall2023}, that the two $z\sim7.66$ sources (IDs 06355, 10612) are closely separated both on the sky ($11''$) and in redshift ($\Delta z=0.005$). Assuming a \citet{Planck2020} cosmology, this corresponds to a projected physical separation and line-of-sight separation of 56~kpc and 179~kpc respectively. Alternatively, the redshift differential translates into a velocity offset of 173~km~s$^{-1}$. We have inspected the lensing model maps and verified that these are two distinct (i.e.\@ not multiply imaged) sources. An investigation into the possible imprint of such a close interaction on the morphologies of these galaxies would require a correction for gravitational lensing, which is beyond the scope of this analysis.

\subsection{Photometric properties} \label{subsec:photometry}

\begin{figure*}
\centering
\includegraphics[width=0.45\linewidth]{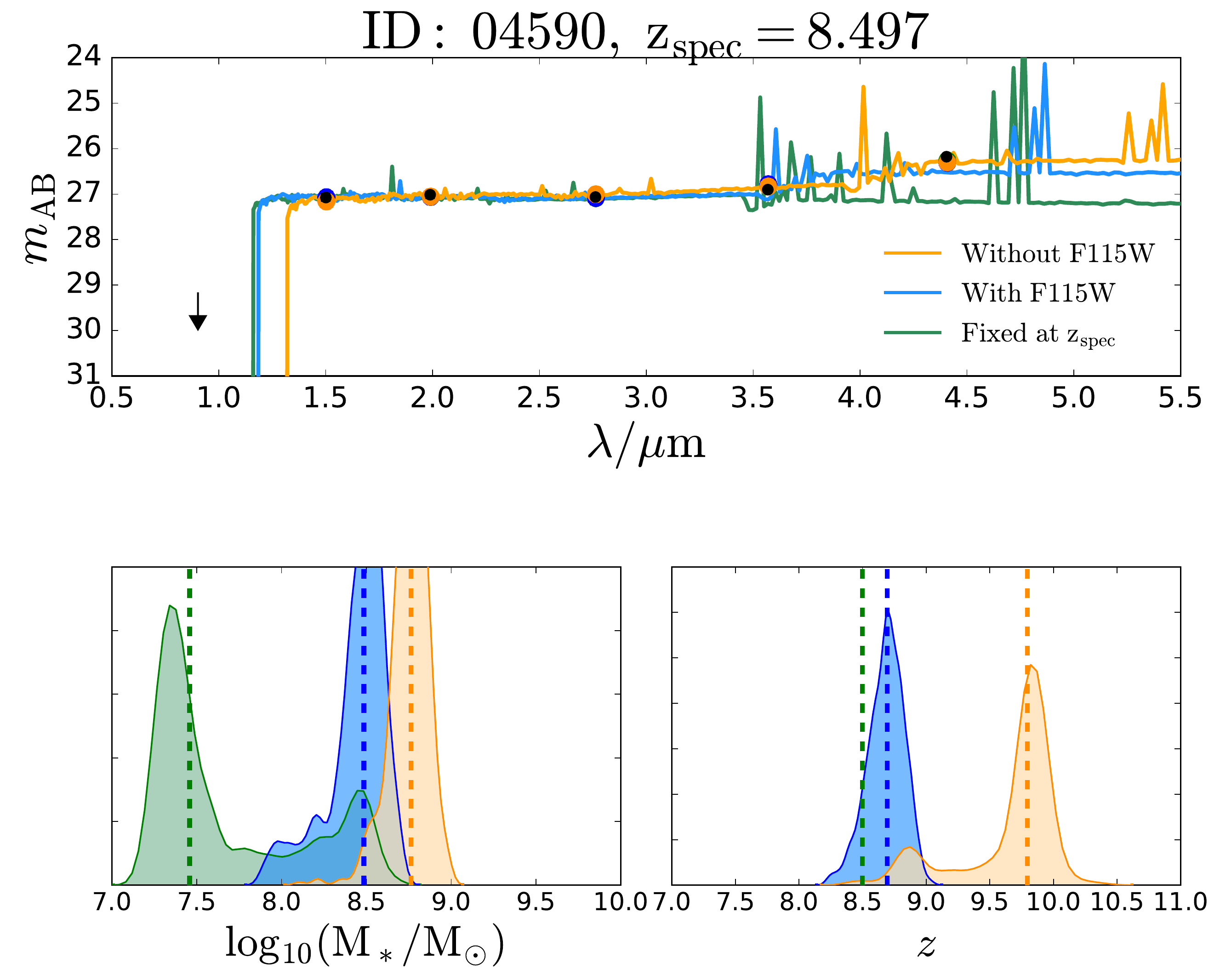} \hfill
\includegraphics[width=0.45\linewidth]{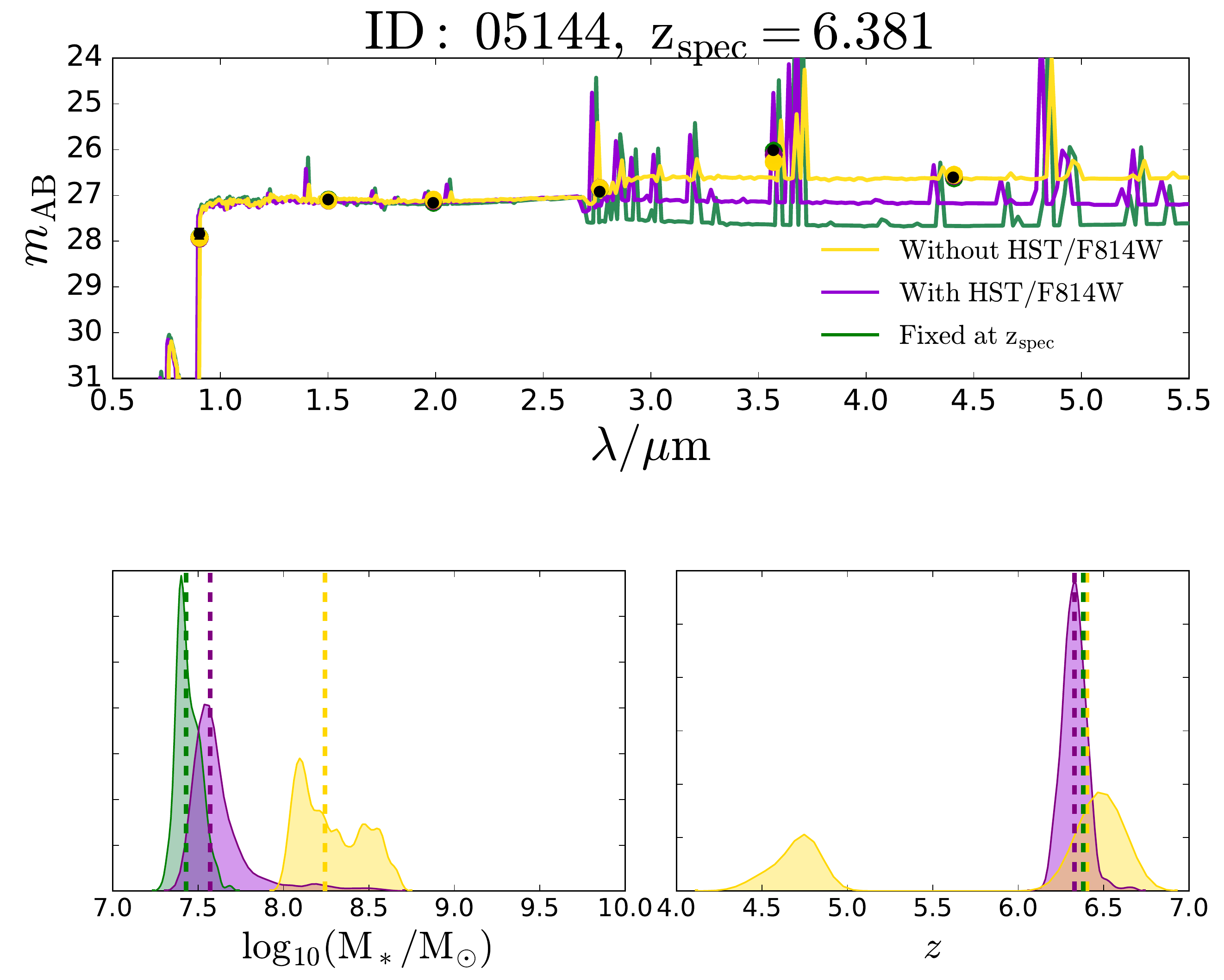} \\[10ex]
\includegraphics[width=0.45\linewidth]{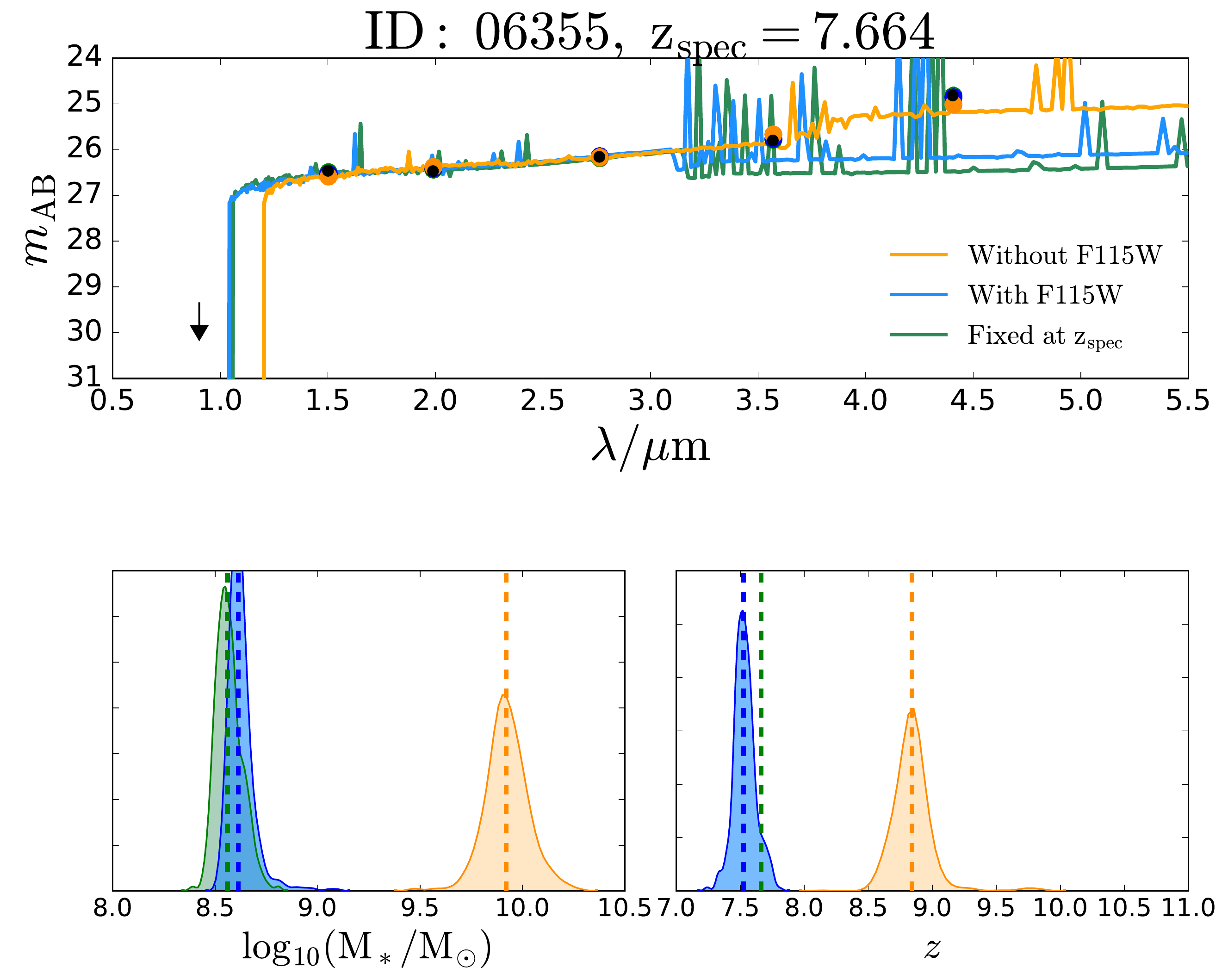}\hfill
\includegraphics[width=0.45\linewidth]{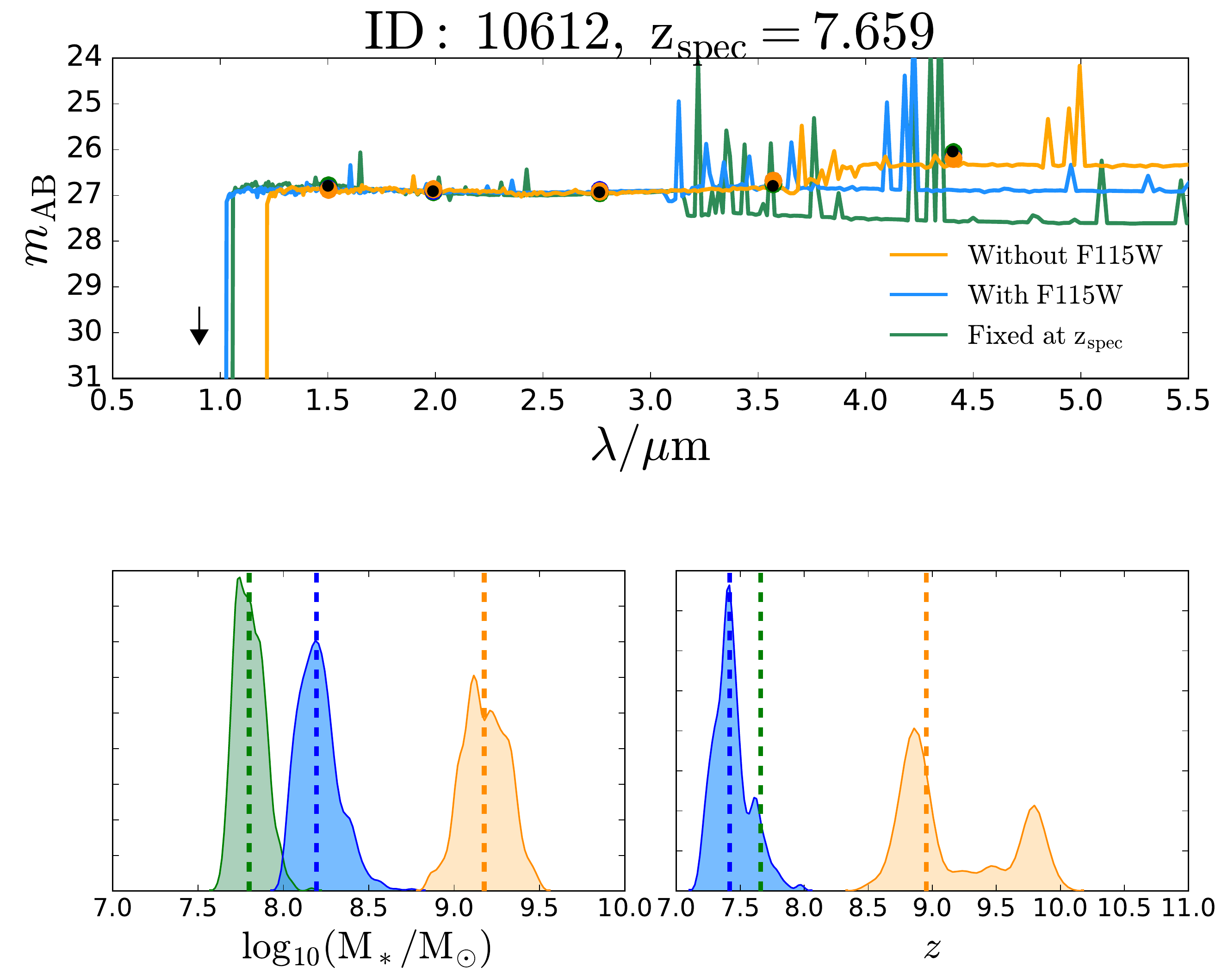}
\caption{Top subpanel: SED fits with BAGPIPES to the available NIRCam photometry (F090W, F150W, F200W, F277W, F356W, F444W, shown in black) for the four EoR galaxies. An exponential star formation history was assumed. Bottom subpanels: Probability distributions for stellar mass $M_*$ (left) and photometric redshift $z$ (right) using the available photometry (i.e.\@ without F115W, orange), including F115W (blue) and keeping the redshift fixed at the known spectroscopic redshift (green). For the $z_\mathrm{spec}=6.381$ galaxy, we instead show the probability distributions without (yellow) and with \emph{HST}/ACS/F814W photometry (purple). The median derived parameters (dashed lines) and spectroscopic redshift (green dashed line) are shown. The mock F115W flux was given by the median BAGPIPES fit obtained when fixing the redshift at the known spectroscopic redshift. Without F115W, the $z_\mathrm{spec}=7.659, 7.664, 8.497$ galaxies are misidentified, with a median assigned photometric redshift $z_\mathrm{phot}\sim 9.5$. This misidentification comes about because the red F356W$-$F444W colour (actually caused by high EW [\ion{O}{III}] $\lambda 5007$ and H$\beta$ emission) is mistakenly attributed to a Balmer break (as shown by the fit in the top subpanel). With the inclusion of F115W, the high-redshift $z\sim10$ solution is ruled out (due to the absence of a Lyman break in F115W), with the photometric redshifts being derived thus being much closer to the true spectroscopic redshift.}
\label{fig:bagpipes_fits}
\end{figure*}

We now turn to the photometry of these four sources. We fit the photometry using the SED fitting code {\tt BAGPIPES} \citet{Carnall2018}, imposing a minimum error of 5\% on the photometry to account for uncertainties on the absolute flux calibration. We assume an exponential star formation history, adopting the default {\tt BAGPIPES} fitting configuration, with uniform priors on the galaxy age (0.001--15~Gyr), $e$-folding time $\tau$ (0.1--10~Gyr), stellar mass formed $\log M_*/\mathrm{M}_\odot$ (1--15), metallicity $Z/\mathrm{Z}_\odot$ (0--2.5), dust extinction $A_\mathrm{V} (0$--$2)$, ionisation parameter $\log U$ ($-4$--$-2$). We assume the \citet{Calzetti2000} dust attenuation curve and use the default spectral templates in {\tt BAGPIPES}.

The measured magnitudes in $0.32''$ diameter extraction apertures (having been corrected using an aperture correction), together with the median {\tt BAGPIPES} fit, as well as the probability distributions for the derived stellar masses (corrected for magnification) and photometric redshifts, are shown in Fig.\@~\ref{fig:bagpipes_fits}. We did not assume the known spectroscopic redshifts for these sources in the fitting process, as we wish to emphasise both the relatively broad range in photometric redshifts that are derived (from the available NIRCam photometry alone, in orange) for these objects, as well as the fact that the $z_\mathrm{spec}=7.659, 7.664, 8.497$ galaxies are misidentified, with a median assigned photometric $z_\mathrm{phot} \sim 9.5$. Thus, the range of possible photometric redshifts derived can deviate quite substantially from the true spectroscopic redshifts (shown by the vertical green dashed line in the redshift subpanel) that are now known for these sources. 

We note that in the NIRSpec MSA target catalog for this ERO program (ID: 2736), the  $z_\mathrm{spec}=7.659, 7.664, 8.497$ galaxies were in fact also assigned photometric redshifts $z_\mathrm{phot}\sim10$. Thus, perhaps the very basis for their spectroscopic follow-up, was that these sources were perceived to be at higher redshifts than their true redshift, i.e.\@ $z_\mathrm{phot} > z_\mathrm{spec}$. Whilst this is certainly the first case in which a misidentified object was selected for spectroscopic follow-up with NIRSpec (being an ERO program), it almost definitely will not be the last.

Indeed, the key spectral features that are largely driving these high-redshift solutions for these sources, are the prominent Lyman-break in the F090W filter, together with the ``magnitude excess'' in the F444W filter \citep[akin to the well-known IRAC excess, see e.g.\@][]{Laporte2014, Smit2014, Smit2015}. However, without the F115W filter (with the corresponding $M_*$ and $z$ probability distributions shown in orange in Fig.\@~\ref{fig:bagpipes_fits}), models are likely to be unable to effectively distinguish between the lower redshift $z\sim8$ galaxies and higher redshift $z\sim10$ galaxies, that can both effectively drive these SED profiles. For the $z\sim8$ galaxies, the excessively bright F444W emission (relative to F356W) is attributable to the high EW [\ion{O}{III}] $\lambda\lambda 4959, \, 5007$ + H$\beta$ emission for these galaxies (which we shall discuss in more detail in Section~\ref{sec:spectroscopy}). For the $z\sim10$ galaxies, the red (F356W$-$F444W) colours are instead likely driven by a prominent Balmer break \citep[see also][]{Wilkins2022b}. With the addition of the F115W filter, one should be able to more effectively distinguish between the two redshift regimes, through the presence or absence of a Lyman break in the F115W filter. Indeed, as shown by the blue probability distributions in Fig.\@~\ref{fig:bagpipes_fits}, by including mock F115W fluxes (as given by the median {\tt BAGPIPES} fit, with the redshift fixed at the known spectroscopic redshift), the higher redshift $z\sim10$ solution is ruled out, and the photometric redshifts derived are much closer to the true spectroscopic redshift.

e note that, in the absence of a F115W measurement, the posterior redshift probability distributions for galaxies 06355 and 10612 are zero at the known spectroscopic redshifts, despite the fact that both $z\sim8$ line emission and $z\sim10$ Balmer breaks should be compatible with the data. The zero probability at the known spectroscopic redshift either indicates that the templates/configuration used in our {\tt BAGPIPES} analysis are unable to produce emission lines of sufficient strength (i.e.\@ high enough equivalent width) to match the observed data (thus yielding zero probability), or that the $z\sim8$ solutions are ''missed'' in the fitting process. We can clearly see  from Fig.\@~\ref{fig:bagpipes_fits} that, upon fixing the redshift in the fitting process to the known spectroscopic redshift, that {\tt BAGPIPES} can well-reproduce (shown in green) the strong F444W flux seen through emission lines alone, achieving excellent agreement with the NIRCam data. Thus the lack of probability at $z\sim8$ when fitting without F115W likely stems from the fact that {\tt BAGPIPES} does not exhaustively fit with every possible redshift and star formation history combination (i.e.\@ a brute-force approach).

From our investigations, this lack of posterior probability at the known spectroscopic redshift occurs rather frequently for strong line emitters in the EoR, even when adopting a variety of SED-fitting codes (such as {\tt Le Phare} and {\tt EAZY}). As discussed more extensively in \citet{Adams2023}, but also in Appendix~\ref{app:photometry}, this is likely due to the spectral templates (that are widely adopted in the use of these codes) not having sufficiently strong emission lines to match the high equivalent width line emission seen in EoR galaxies. Thus, in order to ensure accurate photometric redshifts, it is imperative that templates with strong line emission ([\ion{O}{III}] $\lambda 5007$ EW$_\mathrm{rest}$ $\geq$ 1000~\AA) are adopted when SED-fitting in the \emph{JWST} era.

In order to determine the robustness of our SED-fitting results, we undertake a similar photometric redshift analysis with the SED-fitting codes {\tt Le Phare} and {\tt EAZY} in Appendix~\ref{app:photometry}. Briefly, we obtain the same results as with {\tt BAGPIPES}, namely that the flux excess in the F444W band is attributed to a $z\sim10$ Balmer break, rather than $z\sim8$ [\ion{O}{III}] + H$\beta$ line emission. With the inclusion of mock F115W photometry, both the {\tt Le Phare} and {\tt EAZY} photometric redshifts begin to agree better with the known spectroscopic redshifts.

\subsection{Colour selection}

\begin{figure*}
\centering
\includegraphics[width=\linewidth]{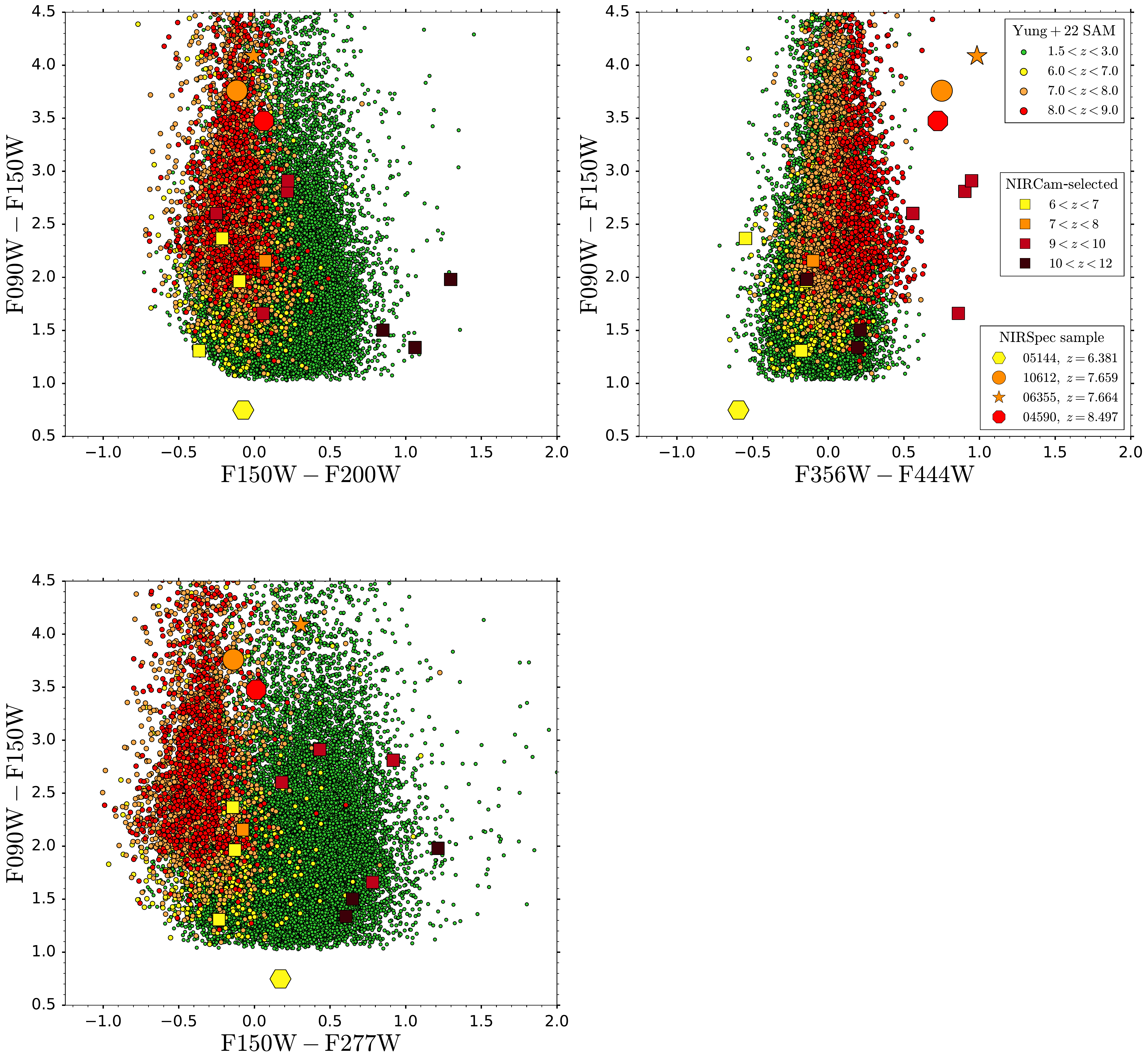}
\caption{Colour--colour selection of galaxies in the epoch of reionisation $(6 < z < 9)$. Top-left: The (F090W$-$F150W), (F150W$-$F200W) colour plane. The basis behind this colour selection is that the (F090W$-$F150W) color probes the Ly$\alpha$ break across this redshift range, which therefore leads to characteristically red colours in this filter pair. The F150W$-$200W colour traces the spectral slope of the continuum. Bottom-left: The (F090W$-$F150W), (F150W$-$F277W) colour plane. The basis behind this colour selection is the same as the top-left panel, but the longer wavelength baseline probed by the (F150W$-$F277W) colour results in a greater separation between the red $z\sim2$ interlopers and the relatively blue/flat EoR galaxies. Right: The (F090W$-$F150W), (F356W$-$F444W) colour plane. Here the red (F356W$-$F444W) colour is tracing the ``magnitude excess'' in the F444W filter (relative to F356W), which is brought about by the high equivalent width [\ion{O}{III}] $\lambda\lambda 4959, 5007$ and H$\beta$  emission in this filter for $z\sim 8$. galaxies. The small points, colour coded by their redshift, are the NIRCam colours predicted by the \citet{Yung2022} semi-analytic model, after introducing noise and selection criteria (F090W non-detection, $\geq 5\sigma$ detections in the other bands) comparable to the observations. The four EoR galaxies studied in this work are shown by the large data points: the $z=6.38$1 galaxy (05144, yellow hexagon), the $z=7.659$ galaxy (10612, orange circle), the $z=7.664$ galaxy (06355, orange star) and the $z=8.497$ galaxy (04590, red octagon). Note that the very red (or blue) F356W-F444W colours exhibited by these four EoR galaxies lie within the extremes predicted by the \citet{Yung2022} semi-analytic model, despite these sources only having been selected over a 5~arcmin${^2}$ area. Our NIRCam-selected sample of high-redshift candidates in the SMACS 0723 field are shown by the colour-coded squares.}
\label{fig:colours_sam}
\end{figure*}

Having discussed the prospects for deriving accurate photometric redshifts from \emph{JWST} photometry, we now focus on the complementary approach of using colour--colour selections to identify high-$z$ candidates. We wish to establish where our known high-redshift objects actually lie in colour--colour space, and whether this is consistent with predictions from theoretical models.

In the top-left panel of Fig.\@~\ref{fig:colours_sam}, we show the (F090W$-$F150W) vs.\@ (F150W$-$F200W) colour plane, which in principle can be used to select $6< z < 10$ galaxies. The basis behind this colour selection is as follows. At these redshifts, the F090W and F150W filters are probing blueward and redward of the Ly$\alpha$ line, respectively. Thus the F090W$-$F150W measures the strength of the Ly$\alpha$ break. Note that this is only the perceived strength of the break, as it is sensitive to the depth of the observations. On the other hand, the F150W$-$F200W colour is sensitive to the spectral slope of the continuum redward of Ly$\alpha$. 

We show the NIRCam colours predicted by an updated version of the \citet{Yung2022} semi-analytic model (which now also includes nebular line emission), with the various coloured data points corresponding to galaxies at different redshifts. We show the positions of $6 < z < 9$ EoR galaxies in the colour--colour plane, together with lower redshift interlopers ($1.5 < z < 3.0$) which can potentially exhibit comparable colours (in this colour plane) to the higher-redshift sources. In order to allow for a comparison with the observational data, we introduced noise into the simulated colours, matching the depths measured by placing $0.32''$ diameter apertures in empty patches of the SMACS 0723 field. Furthermore, we adopt the same selection criteria that would be used for identifying potential high-$z$ candidates, namely $\geq 5\sigma$ detections in the F150W, F200W, F277W, F356W and F444W bands, together with a non-detection in F090W (i.e.\@ with the measured F090W flux being $\leq 2\sigma$). The colours of the four EoR galaxies studied in this work are also shown, with the large octagon, hexagon, star and circle corresponding to the source IDs 04590, 05144, 06355 and 10612, respectively, following the same colour coding as was adopted for the simulated galaxies. Furthermore, we also show (as squares), the colours for our NIRCam-selected high-redshift galaxies in the SMACS 0723 field \citep{Adams2023}. We note that we find no other $8 < z < 9$ candidates, with the red F150W$-$F200W colours for the $10 < z < 12$ candidates being attributable to the fact that these dropout (due to IGM attenuation) in the F150W band.

It is clear from Fig.\@~\ref{fig:colours_sam} that both EoR galaxies and $z\sim2$ galaxies can exhibit similarly red (F090W$-$F150W) colours, due to their Lyman breaks and Balmer breaks, respectively. In principle, these red $z\sim 2$ galaxies have, on average, redder (F150W$-$F200W) colours than EoR sources. However, in practice, owing to finite depth and the fact that the $z\sim2$ population has a much greater density on the sky than the EoR galaxies, contamination can be a considerable concern for selecting targets for follow up.

Importantly, we note that the four EoR galaxies are roughly consistent with the predictions from the \citet{Yung2022} semi-analytic model, with the exact (F090W$-$F150W) colour measured depending on the depth of the observations. It is evident that an application of this colour selection would not have provided a strong basis to select these sources for follow-up spectroscopy with NIRSpec.

In the bottom-left panel of Fig.\@~\ref{fig:colours_sam}, we show the (F090W$-$F150W), (F150W$-$F277W) colour--colour plane. The basis behind this colour selection is similar to that of the top-left panel of Fig.\@~\ref{fig:colours_sam}. However, as the (F150W$-$F277W) colour operates over a longer wavelength baseline than (F150W$-$F200W), it in principle results in even redder colours for $z\sim2$ Balmer break galaxies (due to their red SEDs), and bluer/still flat colours for EoR galaxies. This therefore widens the separation between EoR galaxies and $z\sim2$ contaminants in the colour--colour plane, thus enabling a more effective colour selection. We note that this colour selection was also advocated for in \citet{Harikane2023}. As the \citet{Yung2022} mock catalog is based off a lightcone, the relative number of EoR galaxies and $z\sim2$ galaxies predicted by the semi-analytic model is indicative of what should be seen in observations. Owing to its blue (F150W$-$F277W) colour, there would have been a good basis to identify galaxy 10612 as an EoR candidate. In contrast, the other two $z > 7$ galaxies (04590 and 06355) would not have clearly been identified as EoR candidates from this colour selection alone. 

In the right panel of Fig.\@~\ref{fig:colours_sam}, we show an alternative colour selection that is based on the (F090W$-$F150W) vs.\@ (F356W$-$F444W) colour plane. The basis behind the (F356W$-$F444W) colour selection, traditionally referred to as the IRAC excess, is the high EW [\ion{O}{III}] $\lambda 5007$ and H$\beta$ emission that falls in the F444W filter at $z\sim8$, thus providing a considerable flux boost compared to the neighbouring F356W filter. 

We see that the three $z > 7$ galaxies exhibit relatively very red ($0.6$--$0.9$~mag) (F356W$-$F444W) colours, with the excessively blue (F356W$-$F444W) colour for the $z=6.381$ galaxy being due to the fact that [\ion{O}{III}] $\lambda 5007$ and H$\beta$ instead fall in the F356W filter. What is very evident, is that these four EoR galaxies all have F356W$-$F444W colours that sit on the extreme ends of the \citet{Yung2022} predictions. Now, these high-$z$ galaxies were all observed in a \emph{single} NIRCam module ($\approx5$~arcmin$^2$), while the \citet{Yung2022} catalog covers a field of 782~arcmin$^2$. Thus, barring any observational biases (as it is precisely these ``extreme'' objects that were identified as likely high-$z$ candidates and thus followed up with spectroscopy), this would suggest that the average EoR galaxy likely has stronger line emission (and thus redder F356W$-$F444W colours) than is predicted in the \citet{Yung2022} catalog. Thus, selecting high-$z$ candidates based off of their highly red (F356W$-$F444W) colours (e.g.\@ $\Delta m \geq 0.6$), may not lead to as biased of a galaxy sample as would have otherwise been expected. 

We note that while our 4 EoR galaxies exhibit F090W$-$F150W, F150W$-$F200W and F150W$-$F277W colours that are somewhat consistent with low-$z$ interlopers, we find no low-redshift solutions in our SED-fitting analysis with {\tt BAGPIPES} in Section~\ref{sec:photometry}. Indeed, it is the distinctly red F356W$-$F444W colours of these sources (driven by strong line emission) that helps to break the degeneracy between $z\sim2$ and EoR galaxies.

In Fig.\@~\ref{fig:colours_flares}, we show the same colour--colour selections, but now applied to the colour predictions from the {\footnotesize FLARES} simulations \citep[][]{Lovell2021, Vijayan2021, Vijayan2022, Roper2022, Wilkins2022a, Wilkins2022b}. These simulations are a suite of hydrodynamic simulations of galaxy formation and evolution using the physics of {\footnotesize EAGLE} \citep[][]{Crain2015, Schaye2015a}. The suite consists of a series of 40 zoom resimulations, encompassing a range of overdensities in order to study the environmental effect on high-redshift galaxy evolution. {\footnotesize FLARES} self-consistently models nebular line and continuum emission from young star-forming regions, as well as the effect of the complex 3D geometry of stars and dust on the overall attenuation. Briefly, while the {\footnotesize FLARES} predictions do extend to redder (F356W$-$F444W) colours, we still find that our four EoR galaxies all sit on the extreme ends of the model predictions. 
  
\section{Spectroscopic properties of the EoR galaxies} \label{sec:spectroscopy}

\subsection{Spectral fitting process}

\begin{figure*}
\centering
\includegraphics[width=\linewidth]{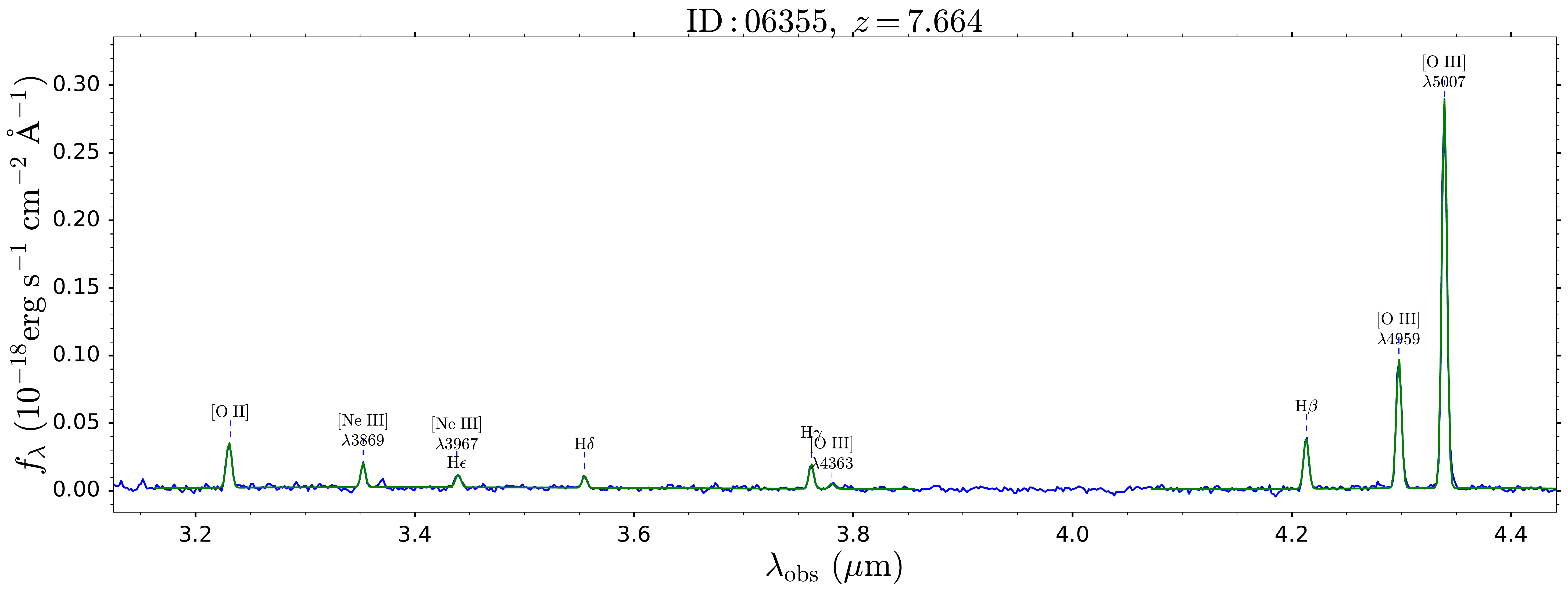}
\caption{The spectrum of the $z=7.664$ galaxy (ID: 06355) using the NIRSpec/G395M grating. The continuum, together with the [\ion{O}{III}] $ \lambda\lambda 4959, 5007$ doublet, the Balmer lines H$\beta$, H$\gamma$, H$\delta$, H$\epsilon$, as well as the [\ion{O}{III}] $\lambda 4363$, [\ion{Ne}{III}] $\lambda\lambda 3869, 3967$ and [\ion{O}{II}] $\lambda\lambda 3726, 3729$ emission lines, have been fit for the four EoR galaxies.}
\label{fig:06355_spectrum}
\end{figure*}

In order to outline our spectral fitting process, we show the spectrum of the $z=7.664$ galaxy (ID: 06355) in Fig.\@~\ref{fig:06355_spectrum}. This spectrum is a striking example of the quality of spectra that NIRSpec will obtain, clearly showing a multitude of emission lines in the rest-frame optical, but now, for the first time, having been obtained for a galaxy deep within the epoch of reionisation. Indeed, as demonstrated in Fig.\@~\ref{fig:06355_spectrum}, we see [\ion{O}{III}] $\lambda\lambda 4959, 5007$ emission, the Balmer lines H$\beta$, H$\gamma$, H$\delta$, H$\epsilon$, as well as [\ion{Ne}{III}] $\lambda\lambda 3869, 3967$, [\ion{O}{II}] $\lambda\lambda 3726, 3729$ and even the characteristically faint emission from the [\ion{O}{III}] $\lambda 4363$ auroral line. Thus, given the incredibly bright emission lines relative to the continuum level (which has not been subtracted) it is evident that the red (F356W$-$F444W) colours, driven by the strong magnitude excess in the F444W filter, are attributable to the high EW emission of [\ion{O}{III}] $\lambda 5007$, rather than a prominent Balmer break in the galaxy spectrum.

Our spectral fitting process is as follows. We begin by co-adding the spectra taken during two separate observations (Observations 007 and 008). We then fit the rest-frame optical emission lines listed above using {\tt LMFIT} \citep{Newville2014}, assuming a Gaussian profile for each line. These Gaussians have been initially broadened according to the spectral resolution $R$ at the observed wavelength of the corresponding emission line. As the spectrum is dominated by emission lines, we begin with a preliminary fit that simultaneously fits all the emission lines as well as the continuum. The derived emission line velocities and widths are then used to define the wavelength ranges spanned by each line ($\pm 5\sigma$ about the line centre). The collective wavelength range spanned by all the emission lines in the spectrum is then masked, and a piecewise fit to the underlying continuum is done. As can be seen from Fig.\@~\ref{fig:06355_spectrum}, we fit the continuum region around the [\ion{O}{III}] $\lambda\lambda 4959, 5007$ doublet and H$\beta$, as well as the continuum spanning the spectral range between the [\ion{O}{II}] doublet and the [\ion{O}{III}] $\lambda 4363$ auroral line. These fits to the continuum are then subtracted from the observed spectrum, and we repeat the initial spectral fitting process, simultaneously fitting all the emission lines in the continuum-subtracted spectrum. We adopted a piecewise continuum fit, as we find that this yields a better characterisation of the variation in the continuum, especially in the vicinity of the emission lines of interest, than can be achieved from parametrically fitting the entire continuum across the full spectral range.

In order to calculate rest-frame equivalent widths for [\ion{O}{III}] $\lambda 5007$ and H$\beta$, we determined the median flux density $f_\lambda$ in the rest-frame spectral range $0.4800 < \lambda_\mathrm{rest} ($\textmu m$) < 0.5050$, after having masked the spectral regions around the H$\beta$, [\ion{O}{III}] $\lambda 4959$ and [\ion{O}{III}] $\lambda 5007$ emission lines, using the $\pm 5\sigma$ line widths established in the spectral fitting process detailed above. We note that the continuum is well-detected for galaxy 06355, so the equivalent width derived should be reliable. For galaxies 04590, 05144 and 10612, the continuum is instead noisier, with the median flux density estimate being rather sensitive to both the exact wavelength range that is used in the continuum estimate, and the quality of the background subtraction that has been performed in this range. Thus we note that the equivalent widths derived for these sources are less robust. 

Finally, in order to generate errors on emission line fluxes and the continuum level, we generated 100 realisations of the observed spectrum, each perturbed according to the noise level in the observations, and subsequently fit using the procedure outlined above. 

\subsection{Emission line diagnostics}

Having fit the spectrum and derived emission line fluxes, we now compute standard emission line diagnostics used at lower redshifts, which give insights into the ISM conditions within these high-redshift galaxies.

\subsubsection{Measurement method}

As had been shown in Fig.\@~\ref{fig:eor_cutouts}, it is evident that the four EoR galaxies studied in this work exhibit an extended and complex structure. They are therefore not well-described as point sources, nor as uniform objects. Hence the two default source morphologies that can be assumed when correcting for slit losses in the \emph{JWST}/NIRSpec reduction pipeline, will likely not yield an accurate flux calibration for these galaxies. Since this flux calibration is wavelength-dependent, owing to the increase in diffraction (and thus slit loss) with increasing wavelength, this renders a direct comparison between the fluxes between lines at widely separated wavelengths challenging, as the line ratios measured will depend on both the intrinsic line ratio and the differential flux calibration between the two lines. An accurate flux calibration as a function of wavelength will therefore require an accurate assessment of the actual light profile of the galaxy (as indicated by photometry) and the exact position of the galaxy within the NIRSpec slit. Given the detailed, but likely necessary analysis that would have to be undertaken to achieve this, we instead adopt a different approach within this work.

We have chosen to be conservative, by only considering emission line ratios between lines of comparable wavelength, as for these the relative flux calibration should be approximately the same. Hence we focus our analysis on both the [\ion{O}{III}] $\lambda 5007$/H$\beta$ line ratio, as well as the [\ion{Ne}{III}] $\lambda 3869$ /[\ion{O}{II}] $\lambda\lambda 3726, 3729$ line ratios. Furthermore, we also determine the [\ion{O}{III}] $\lambda 5007$ and H$\beta$ rest-frame equivalent widths, as these should also be reliable, provided that the background subtraction is accurate and the spectra are of sufficient depth to effectively measure the continuum level. 

The aforementioned diagnostics can place constraints on the ISM, with the [\ion{O}{III}] $\lambda 5007$/H$\beta$ and [\ion{Ne}{III}] $\lambda 3869$ /[\ion{O}{II}] $\lambda\lambda 3726, 3729$ line ratios being effective tracers of the ionisation parameter $U$ \citep[see e.g.\@][]{Witstok2021}, while the [\ion{O}{III}] $\lambda 5007$ and H$\beta$ EWs are tracers of young, vigorously star-forming (in terms of sSFR) systems \citep[see e.g.\@][]{Belfiore2018b}.

\begin{figure*}
\centering
\includegraphics[width=\linewidth]{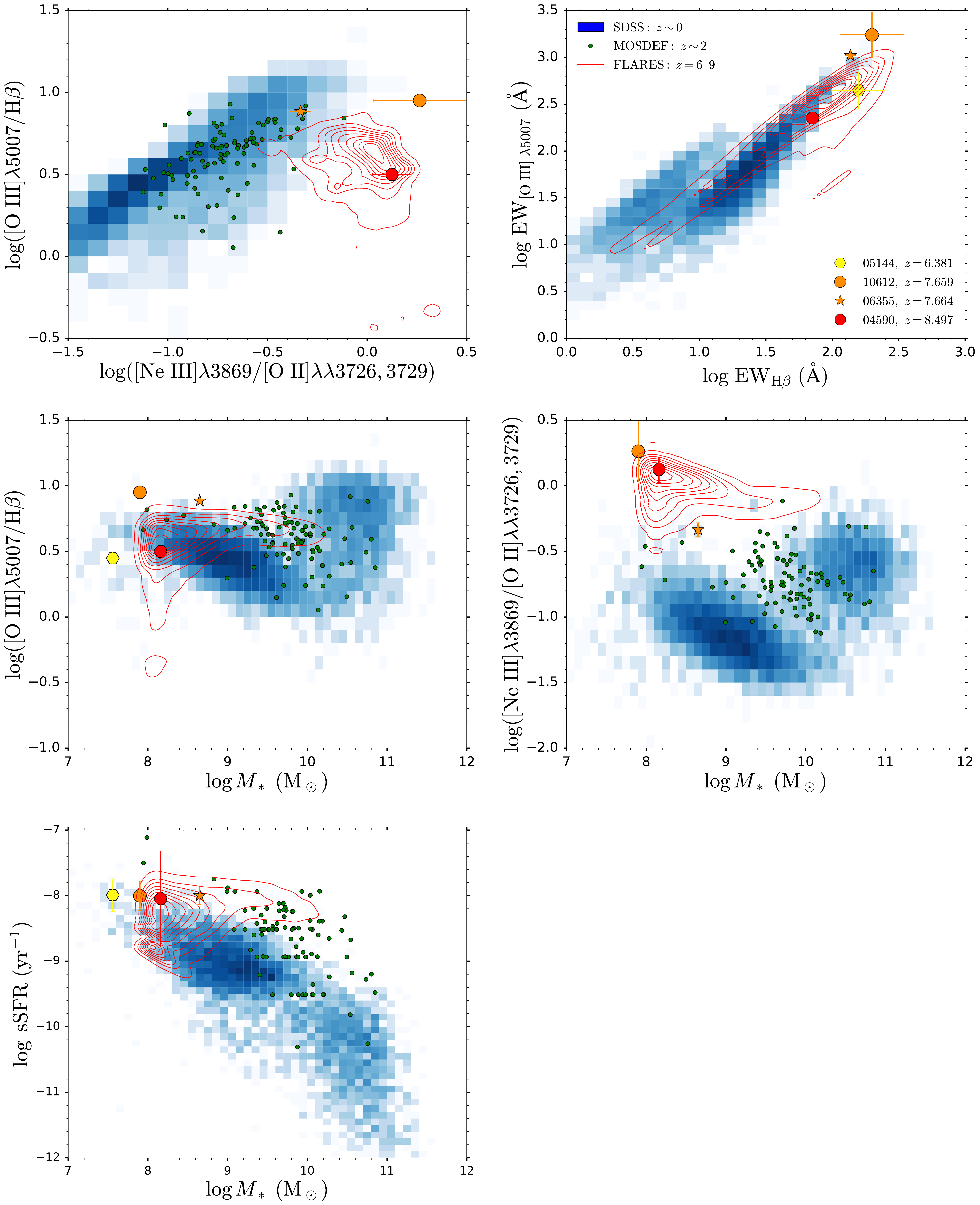}
\caption{Spectroscopic diagnostic diagrams for assessing the nebular conditions within galaxies. We show the properties of $z\sim 0$ SDSS galaxies (blue), $z\sim2$ MOSDEF galaxies (green) and the predictions from the FLARES simulations at $z=6,\ 7,\ 8,\ 9$ (red). Overlaid are the properties of the four sources studied in this work (when the relevant emission lines have been detected and fit in the spectrum), following the same colour coding and symbols used in Fig.\@~\ref{fig:colours_sam}. Top-left: The [\ion{O}{III}] $\lambda 5007$/H$\beta$ –  [\ion{Ne}{III}] $\lambda 3869$ /[\ion{O}{II}] line ratio diagram. We find that the EoR galaxies likely deviate (by $\approx$ 0.5~dex) from the local correlation, having enhanced [\ion{Ne}{III}] $\lambda 3869$ /[\ion{O}{II}] ratios, in line with the FLARES predictions. Top-right: The rest-frame [\ion{O}{III}] $\lambda 5007$ and H$\beta$ equivalent widths. The four EoR galaxies exhibit very high EW [O III] $\lambda 5007$ and H$\beta$ emission (which is consistent with their extreme F356W$-$F444W colours). Whilst this is reserved to extreme, vigorously star-forming dwarf galaxies in the local Universe, our data and the FLARES predictions indicate that it is much more common in the epoch of reionisation. Middle: [\ion{O}{III}] $\lambda 5007$/H$\beta$ (left) and [\ion{Ne}{III}] $\lambda 3869$ /[\ion{O}{II}] as a function of stellar mass. It is primarily the [\ion{Ne}{III}] $\lambda 3869$ /[\ion{O}{II}] emission that is offset from the local trends. Bottom-left: sSFR against stellar mass.}

\label{fig:spectroscopic_diagnostics}
\end{figure*}

We show these spectral diagnostics in Fig.\@~\ref{fig:spectroscopic_diagnostics}. The locus of points occupied by $0.02 < z < 0.085$, $\log (M_*/\mathrm{M}_\odot) > 7$ SDSS DR7 galaxies \citep{Kauffmann2003, Brinchmann2004, Abazajian2009} is shown in blue, the green data points correspond to the $z\sim2$ galaxies from the MOSDEF survey \citep[a sample of $\sim$1500 galaxies and AGN at $1.37 < z < 3.8$ observed with Keck/MOSFIRE, see][]{Kriek2015, Reddy2015}, and the red distribution corresponds to the predictions from the {\footnotesize FLARES} simulations at $z=6$--$9$. Overlaid are shown the measured line ratios and equivalent widths for the four EoR galaxies. For the SDSS galaxies, we applied a S/N cut of 5 to both the measured line fluxes and the continuum flux densities, to ensure that the line fluxes and equivalent widths we show are reliable. For the MOSDEF galaxies, we relaxed these requirements somewhat, requiring a S/N > 5 for [\ion{O}{III}] $\lambda 5007$ and a S/N > 3 for the remaining emission lines. We note that applying such S/N cuts likely introduces biases into our comparison sample, as sources with weak emission lines are preferentially removed. We restrict our analysis to {\footnotesize FLARES} galaxies with $\log (M_*/\mathrm{M}_\odot) > 8$, as below this mass regime, one starts to approach the gas mass resolution of the simulation $\log (M_\mathrm{g}/\mathrm{M}_\odot) = 6$. 

\subsubsection{Revealed emission line properties}

In the top-left panel of Fig.\@~\ref{fig:spectroscopic_diagnostics}, we show the [\ion{O}{III}] $\lambda 5007$/H$\beta$ and [\ion{Ne}{III}] $\lambda 3869$ /[\ion{O}{II}] line ratio diagram. We see that these two line ratios are clearly correlated for local SDSS galaxies, with these galaxies exhibiting a rather tight sequence within this plane, with [\ion{O}{III}] $\lambda 5007$/H$\beta$ tending to increase with increasing [\ion{Ne}{III}] $\lambda 3869$ /[\ion{O}{II}]. Furthermore, the $z\sim 2$ MOSDEF galaxies also tend to follow this local correlation. However, it is clear, even with our limited sample size of three EoR galaxies (for which [\ion{Ne}{III}] $\lambda 3869$ and [\ion{O}{II}] have been detected), that galaxies in the epoch of reionisation likely deviate (by $\approx$ 0.5~dex) from the local trend. Indeed, this qualitative departure from the local correlation is in line with the {\footnotesize FLARES} simulations, who find that [\ion{Ne}{III}] $\lambda 3869$ /[\ion{O}{II}] tends to be higher for a given [\ion{O}{III}] $\lambda 5007$/H$\beta$ for galaxies at $z=7$--$9$ than at $z=0$. However, we note that only one of the galaxies (04590) has line ratios that are quantitatively consistent within the {\footnotesize FLARES} predictions. These enhanced [\ion{Ne}{III}] $\lambda 3869$ /[\ion{O}{II}] ratios in the SMACS 0723 EoR galaxies were also seen in the study of \citet{Trump2022}, which attributed this to these galaxies likely having higher ionisation ($\log Q \approx 8$, i.e.\@ $\log U \approx -2.5$), lower metallicity ($Z \lessapprox 0.1~\mathrm{Z}_\odot$) and higher pressure ($\log P(k) \approx 8$--$9$) than the $z\sim 0$ galaxies. As the {\footnotesize FLARES} simulations assume an ionisation parameter of $\log U \sim -2$, their enhanced  [\ion{Ne}{III}] $\lambda 3869$ /[\ion{O}{II}] ratios are likely driven by the lower metallicity in galaxies at $z=6$--$9$.

Beyond looking at line ratio correlations, it is worthwhile to consider how line ratios depend on stellar mass \citep[through so-called mass–excitation diagrams,][]{Juneau2011} as this can give insights into the mass regime within which different nebular excitation processes (such as photoionisation from star formation vs.\@ AGN) begin to dominate \citep[see e.g.\@][]{Baldwin1981, Kewley2001} Indeed, as shown in the bottom-left panel, SDSS galaxies clearly exhibit a bimodality in the [\ion{O}{III}] $\lambda 5007$/H$\beta$–$M_*$ plane. On the one hand, we have the star-forming main sequence, for which the [\ion{O}{III}] $\lambda 5007$/H$\beta$ ratio tends to decrease with increasing stellar mass. On the other hand, in the high mass regime ($\log (M_*/\mathrm{M}_\odot) > 10$), we begin to see elevated [\ion{O}{III}] $\lambda 5007$/H$\beta$ line ratios, as AGN activity starts to dominate \citep[see e.g.\@][]{Juneau2011, Stanway2014, Dickey2016}. Just as the $z\sim 2$ MOSDEF galaxies tend to have elevated [\ion{O}{III}] $\lambda 5007$/H$\beta$ line ratios at a given stellar mass \citep[with masses from][]{Duncan2014, Duncan2019} compared to the local population (which is attributed to $z\sim2$ galaxies having higher sSFR at a given stellar mass than galaxies at $z\sim0$, \citealt{Dickey2016}, and/or lower nebular metallicity and higher ionisation parameter, \citealt{Sanders2018}), we find that two of our sources (06355 and 10612) exhibit line ratios that are on the extreme end of the local trends. The other two objects (04590 and 05144) have comparable [\ion{O}{III}] $\lambda 5007$/H$\beta$ line ratios to $z=0$ dwarf galaxies. These results are entirely consistent with the (extreme ends) of the {\footnotesize FLARES} predictions, which do not include the AGN contribution to these lines. Hence the [\ion{O}{III}] $\lambda 5007$/H$\beta$ emission in these four galaxies is consistent with what is seen in what is considered extreme, vigorously star-forming dwarf galaxies in the local Universe \citep[see e.g.\@][]{Izotov2021a}.

Similarly, in the bottom-right panel of Fig.\@~\ref{fig:spectroscopic_diagnostics}, we show that SDSS galaxies also exhibit a bimodality in the [\ion{Ne}{III}] $\lambda 3869$/[O II]–$M_*$ plane, which is once again attributable to the relative roles of star formation and AGN activity in the low- and high-mass regimes. As discussed extensively in \citet{Jeong2020}, who analysed stacked measurements of a representative sample of $z\sim2$ MOSDEF star-forming galaxies, as opposed to a biased subset (which we show here) with individual [\ion{Ne}{III}] detections, MOSDEF galaxies have enhanced [\ion{Ne}{III}] $\lambda 3869$/[O II] relative to the SDSS population. They find that the [\ion{Ne}{III}] $\lambda 3869$/[O II], [\ion{O}{III}] $\lambda 5007$/H$\beta$, [\ion{O}{III}]/[\ion{O}{II}] and [\ion{Ne}{III}]/[\ion{O}{II}] ratios for stellar-mass-stacked galaxies have a small but significant offset relative to the median sequence of local SDSS star-forming galaxies. This offset is interpreted as being attributable to an enhanced $\alpha$/Fe ratio (i.e.\@ a harder ionising spectrum) at fixed gas-phase oxygen abundance in star-forming galaxies at higher redshift. Whilst the $z\sim2$ MOSDEF galaxies exhibit slightly enhanced [\ion{Ne}{III}] $\lambda 3869$/[O II] line ratios compared to the local population, it is evident that there is a dramatic evolution in this line ratio as we enter the epoch of reionisation. Indeed, these line ratios begin to exceed what is even seen in local AGN, and is consistent with the {\footnotesize FLARES} predictions for these galaxies. Thus the offset from the local [\ion{O}{III}] $\lambda 5007$/H$\beta$ – [\ion{Ne}{III}] $\lambda 3869$ /[\ion{O}{II}] correlation is attributable to the greatly enhanced  [\ion{Ne}{III}] $\lambda 3869$ /[\ion{O}{II}] line ratios seen in these systems, rather than any significant evolution in [\ion{O}{III}] $\lambda 5007$/H$\beta$.

In order to investigate to what degree the aforementioned trends in the $\lambda 5007$/H$\beta$ and [\ion{Ne}{III}] $\lambda 3869$ /[\ion{O}{II}] line ratios are tied to the redshift-evolution of galaxy sSFRs, we show the sSFR--$M_*$ plane in the bottom-left panel of Fig.\@~\ref{fig:spectroscopic_diagnostics}. We note that the SFRs for SDSS galaxies are H$\alpha$-based \citep{Brinchmann2004}, the SFRs for MOSDEF galaxies \citep{Duncan2014, Duncan2019} and the SMACS 0723 galaxies are SED-based and the {\footnotesize FLARES} SFRs are averaged over the past 10~Myr. We find that the sSFRs of our EoR galaxies lie at the extreme upper ends of the sample of $z\sim0$ SDSS galaxies and are consistent with the predictions from the {\footnotesize FLARES} simulations. The enhanced sSFRs of these EoR galaxies are therefore qualitatively consistent with the higher sSFRs (at a fixed stellar mass) seen in the MOSDEF galaxies. Despite these elevated sSFRs however, our sample of EoR galaxies exhibit $\lambda 5007$/H$\beta$ line ratios that are, on average, consistent with the $z\sim0$ SDSS galaxies. Additionally, owing to the fact that the measured [\ion{Ne}{III}] $\lambda 3869$ /[\ion{O}{II}] ratios for our EoR galaxies extend well beyond ($\sim$ 0.5~dex) the SDSS and MOSDEF trends (for a given stellar mass), while their sSFRs are (roughly) comparable, this suggests that these enhanced [\ion{Ne}{III}] $\lambda 3869$ /[\ion{O}{II}] ratios cannot be due to elevated sSFRs alone.

Finally, in the top-right panel, we show the correlation between the rest-frame [\ion{O}{III}] $\lambda 5007$ and H$\beta$ equivalent widths. It is immediately evident, from comparing the trends for SDSS and {\footnotesize FLARES} galaxies, that very high EW [\ion{O}{III}] $\lambda 5007$ and H$\beta$ emission, which is reserved to extreme, vigorously star-forming dwarf galaxies in the local Universe \citep[see e.g.\@][]{Izotov2021a}, is much more common in the epoch of reionisation. Indeed, our four EoR galaxies roughly fall on the extreme boundaries of the SDSS population, and close to the {\footnotesize FLARES} predictions. Still, as can be seen from Fig.\@~\ref{fig:spectroscopic_diagnostics}, and is consistent with our earlier discussion of the extremely red (F356W$-$F444W) ($>0.6$~mag) colours for the $z > 7$ EoR galaxies (caused by intense [O III] $\lambda 5007$ emission), the [\ion{O}{III}] $\lambda 5007$ equivalent widths for two of our objects (06355, 10612) sit on the extreme ends of the {\footnotesize FLARES} predictions, despite the fact that these objects were only selected from 5~arcmin$^2$ of NIRCam imaging, indicating that this is the norm (or at least, more common) at $z>7$ rather than being the exception.

Indeed, the measured F356W$-$F444W colours from the photometry can be compared against the [\ion{O}{III}] $\lambda 5007$ EW measurements made using spectroscopy. Assuming an otherwise flat $f_\nu$ spectrum (i.e.\@ assuming $\beta = -2$), the magnitude excess $\Delta m$ in the F444W filter (with respect to F356W) is roughly given by $\Delta m = 2.5\log (1 + \mathrm{EW_{obs, total}}/\Delta\lambda)$, where $\mathrm{EW_{obs,total}}$ is the total observed-frame equivalent width of all the emission lines (essentially [\ion{O}{III}] $\lambda\lambda 4959, 5007$ and H$\beta$ at these redshifts) in the F444W filter, and $\Delta\lambda$ is the bandpass width of the F444W filter ($\approx 1$~\textmu m = 10000~\AA). Now, the rest-frame equivalent width $\mathrm{EW_{rest}} = \mathrm{EW_{obs}}/(1+z)$. Thus, given the fixed line ratio between [\ion{O}{III}] $\lambda 4959$ and [\ion{O}{III}] $\lambda 5007$, and assuming the measured [\ion{O}{III}] $\lambda 5007$/H$\beta$ ratio from the observations, one can approximately estimate the [\ion{O}{III}] $\lambda 5007$ rest-frame equivalent width implied by the measured F356W$-$F444W colours. We find colour-based rest-frame equivalent widths of 613~\AA, 474~\AA, 1181~\AA\ and 812~\AA, for galaxies 04590, 05144, 06355 and 10612, respectively. These are indeed roughly comparable (to within 0.45~dex, though see the following discussion) to our estimates from spectroscopy, namely 225~\AA, 444~\AA, 1050~\AA\ and 1740~\AA. Furthermore, our directly measured and inferred [\ion{O}{III}] $\lambda 5007$ equivalent widths are consistent with those inferred from the IRAC excess seen in EoR galaxies \citep[see e.g.\@][]{Smit2014, Endsley2021}.

Any potential discrepancy between the colour-based and spectroscopic [\ion{O}{III}] $\lambda 5007$ EW estimates are primarily due to three factors. Firstly, from the assumption of a flat $f_\nu$ continuum when inferring the [\ion{O}{III}] $\lambda 5007$ contribution to the magnitude excess in the F444W filter. Indeed, galaxy 06355 clearly has a red continuum (see Fig.\@~\ref{fig:bagpipes_fits}, causing our colour-based EW measurement (613~\AA) to overestimate the spectroscopic value (225~\AA). Secondly, due to the difficulty in estimating the continuum level from spectroscopy. As mentioned earlier, while the continuum is clearly detected for 06355, the derived measurements are more sensitive to the exact wavelength range adopted for continuum estimation for galaxies 04590, 05144 and 10612. Indeed, with deeper data, and an improved background subtraction (building on the current default reduction pipeline, which is currently affected by snowballs from cosmic rays), more reliable spectroscopy-based [\ion{O}{III}] $\lambda 5007$ EW estimates for galaxies in the epoch of reionisation will be possible. Owing to the noisy continuum for galaxy 10612, its spectroscopy-based EW measurement ($1740\pm976$~\AA) is still consistent with the smaller colour-based value (812~\AA). Thirdly, the photometry and spectroscopy probe different regions within the galaxies, and thus likely different average nebular conditions.

\subsection{Distinguishing between star formation and AGN}

\begin{figure*}
\centering
\includegraphics[width=\linewidth]{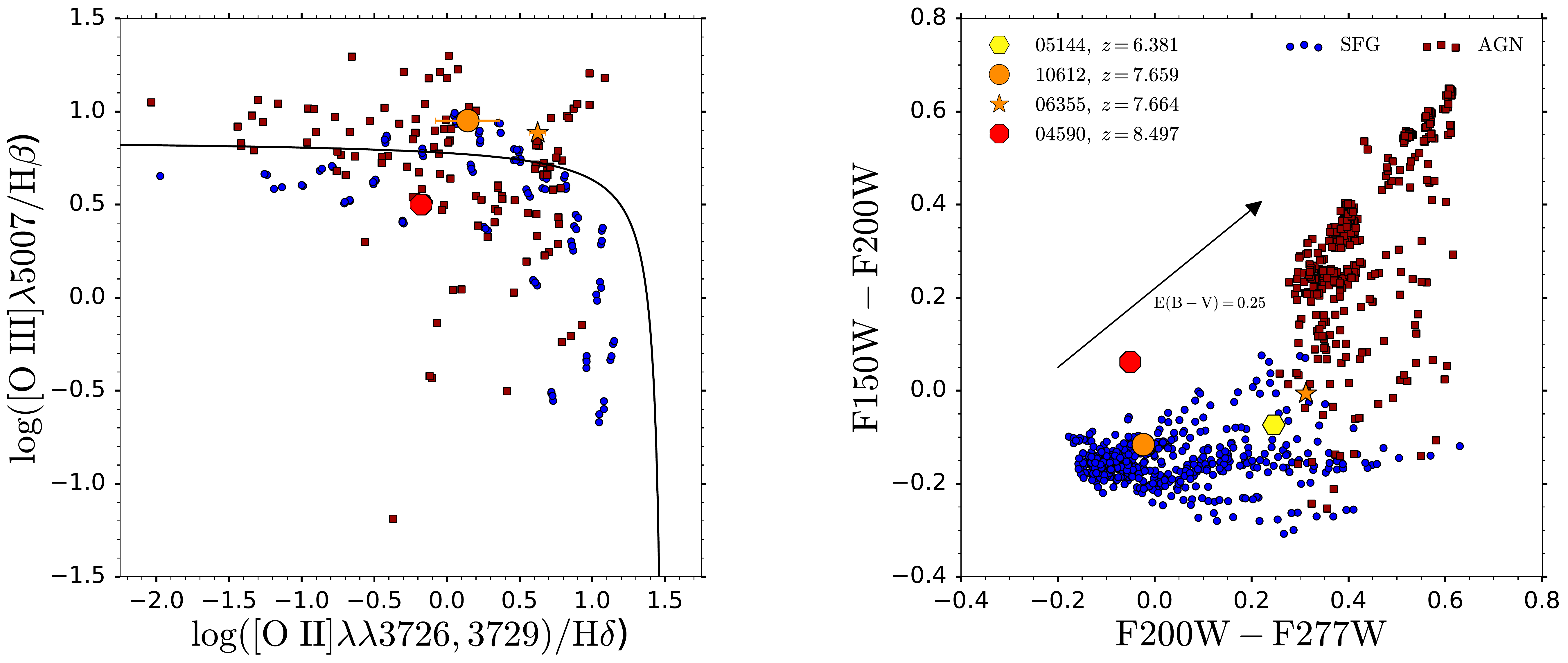}
\caption{Spectroscopic and photometric diagnostics for separating star-forming galaxies and AGN in the epoch of reionisation. Left: The [\ion{O}{III}] $\lambda 5007$/H$\beta$–[\ion{O}{II}]/H$\delta$ BPT diagram, the lines for which are all accessible by NIRSpec at $z < 9.6$. We show the \citet{Lamareille2010} boundary separating $z=0$ SDSS star-forming galaxies and AGN (black line), as well as the line ratios predicted by the star-forming galaxy (blue circles) and AGN (dark red squares) models by \citet{Nakajima2022}. Given the degree of overlap between star-forming galaxies and AGN, it is not possible to make a definitive assessment of the relative roles of star formation and AGN in driving the nebular emission in the EoR galaxies.  Right: The F150W$-$F200W, F200W$-$F277W colour--colour plane. The basis behind this colour selection is the (assumed) power law continuum emission in AGN galaxies, which results in distinct, red colours in the F150W$-$F200W, F200W$-$F277W filter pairs, which probe the rest-frame UV at the redshifts shown ($z=7,\ 8,\ 9$). As indicated by the reddening vector, this rest-frame UV light can be strongly affected by dust reddening (which is not included in our implementation of the \citealt{Nakajima2022} models), potentially causing star-forming galaxies to overlap with AGN in this plane. Thus through the combination of photometry (to identify AGN candidates) and spectroscopy (to rule out low-redshift interlopers and to correct for dust reddening), it should in principle be possible to separate star-forming galaxies and AGN in the EoR.}
\label{fig:bpt}
\end{figure*}

Given the elevated [\ion{Ne}{III}] $\lambda 3869$ /[\ion{O}{II}] ratios in EoR galaxies, which are somewhat comparable to those seen in local AGN, we now discuss the prospects of distinguishing between nebular emission primarily driven by star formation and AGN activity in the epoch of reionisation. 

Traditionally, rest-frame optical spectroscopy can be used to separate star-forming galaxies and AGN through the application of BPT diagrams \citep[see e.g.\@][]{Baldwin1981, Kewley2001, Steidel2014, Shapley2015a, Belfiore2016, Curti2022a} Indeed, through comparing the line ratios for doubly-ionised species, such as [\ion{O}{III}] $\lambda 5007$/H$\beta$ against the line ratios for singly-ionised or neutral species, such as [\ion{N}{II}]/H$\alpha$, [\ion{S}{II}]/H$\alpha$ or [\ion{O}{I}]/H$\alpha$, one can typically identify the primary nebular excitation source in galaxies. However, as pointed out in \citet{Nakajima2022}, at $z > 7$, the H$\alpha$, [\ion{N}{II}], [\ion{S}{II}] and [\ion{O}{I}] lines get redshifted out of the spectral range of NIRSpec and into the spectral range of \emph{JWST}/MIRI. Given MIRI's lower sensitivity (roughly $6\times$ less than NIRSpec/G395M) and lack of multiplexing, acquiring large statistical spectroscopic samples of galaxies will be a challenging endeavour, which will likely be gradually built up over \emph{JWST}'s lifetime. Thus, alternative emission line ratios, which are accessible to NIRSpec at these redshifts, are necessary.

\citet{Lamareille2010} showed that the [\ion{O}{III}] $\lambda 5007$/H$\beta$–[\ion{O}{II}]/H$\beta$ diagram is a possible alternative, which can be applied to help separate star-forming galaxies from AGN. Indeed, \citet{Nakajima2022} have recently further investigated the applicability of this diagnostic, using their new spectral models for metal-poor Pop II galaxies ($Z_\mathrm{gas} \leq 0.1~\mathrm{Z}_\odot)$, metal-enriched Pop I galaxies ($Z_\mathrm{gas} \geq 0.1~\mathrm{Z}_\odot)$ and AGN (as well as Pop III galaxies and direct collapse black holes, which we do not explore here).

The \citet{Lamareille2010} boundary, which separates $z=0$ SDSS star-forming galaxies and AGN, is shown as the solid line in the left panel of Fig.\@~\ref{fig:bpt}. The \citet{Nakajima2022} models for star-forming galaxies (Pop I and Pop II) are shown in blue, while their models for AGN are shown in dark red. The line ratios for the three EoR galaxies (which have the relevant line detections) are also shown. Note that we show here the [\ion{O}{II}]/H$\gamma$ ratio, rather than the traditional [\ion{O}{II}]/H$\beta$ ratio. For the reasons mentioned earlier, we have decided to adopt a conservative approach, where uncertainties in the flux calibration as a function of wavelength have been (attempted to be) mitigated, by computing ratios between two lines of comparable wavelength. The \citet{Lamareille2010} boundary was thus adapted by assuming an intrinsic H$\gamma$/H$\beta$ line ratio of 0.259 (i.e.\@ ignoring reddening by dust).

As can be seen from Fig.\@~\ref{fig:bpt}, the \citet{Nakajima2022} line ratios for AGN overlap to some degree with the models for star-forming galaxies. As pointed out in their paper, however, AGN can be isolated provided that their metallicity is relatively high ($Z  \gtrsim 0.5~\mathrm{Z}_\odot$) and the spectrum is hard (with the power law slope $\alpha \gtrsim -1.6$). These are precisely the AGN data points that exist in the upper regions of the [\ion{O}{III}] $\lambda 5007$/H$\beta$–[\ion{O}{II}]/H$\gamma$ diagram. As can be seen from Fig.\@~\ref{fig:bpt}, it would in principle not be possible to make a definitive assessment of the relative roles of star-formation and AGN in driving the nebular emission in these three EoR galaxies.

We thus highlight the prospects for using photometry to distinguish between star-forming galaxies (SFG) and AGN in the epoch of reionisation. Indeed, it is the synergy between photometry and spectroscopy that could potentially aid in such identifications. Assuming the continuum emission in AGN follows a power law, with $f_\nu \propto \nu ^{-\alpha}$, the colours for AGN will be very red, especially in the rest-frame UV where the power law is steepest in the \citet{Nakajima2022} models. Hence in the epoch of reionisation, where the UV continuum redward of Ly$\alpha$ sits in the F150W, F200W and F277W NIRCam bands, one would expect to find characteristically red colours in the (F150W$-$F200W) and (F200W$-$F277W) filter pairs. 

Indeed, as shown in the right panel of Fig.\@~\ref{fig:bpt}, the \citet{Nakajima2022} models for AGN spectra (which follow a power law continuum) in the redshift range $7 < z < 9$ exhibit distinctly red colours, which clearly separates them from star-forming galaxies at the same redshifts. Whilst such red colours could potentially be attributed to lower-redshift interlopers, through the addition of spectroscopy and a measurement of the spectroscopic redshift, such interlopers can be definitively ruled out. We do note that the rest-frame UV light can be strongly affected by dust reddening, which is not included in our implementation of the \citet{Nakajima2022} models shown in Fig.\@~\ref{fig:bpt}. Indeed, as shown by the reddening vector in the right panel, such dust reddening can readily cause star-forming galaxies to overlap with AGN in the F150W$-$F200W, F200W$-$F277W colour--colour plane, making an AGN/SFG identification more difficult from just a colour selection alone. This thus highlights an additional benefit of spectroscopy, as through a measurement of the Balmer decrement, through e.g.\@ H$\beta$/H$\gamma$ (which are both accessible by NIRSpec at these redshifts), one can constrain the amount of dust $\mathrm{E(B-V)}$ and thus determine the dust-corrected colours. This should therefore allow for a more effective identification of star-forming galaxies and AGN in the F150W$-$F200W, F200W$-$F277W colour--colour plane. Hence through the combination of photometry and spectroscopy it should in principle be possible to separate star-forming galaxies and AGN in the EoR.

From looking at the positions of our four EoR galaxies in the F150W$-$F200W, F200W$-$F277W plane, we see their colours are largely consistent with what one would expect for star-forming galaxies. A more thorough investigation, utilising a more diverse set of star-forming and AGN models to analyse both the source colours and their full photometry through SED fitting, would place more stringent constraints on the relative roles of star-formation and AGN activity in galaxies (such as those studied here) in the epoch of reionisation.

Finally, we note that rest-frame UV spectroscopy will also aid in distinguishing between star-forming galaxies and AGN in the EoR. As discussed and investigated extensively in \citet{Brinchmann2022}, rest-frame UV emission lines such as \ion{C}{III}] $\lambda\lambda 1907,1909$ and [\ion{Ne}{IV}] $\lambda\lambda 2422, 2424$ can serve as valuable indicators of potential AGN activity, with these rest-UV lines all being accessible by NIRSpec at these redshifts.

\section{Summary and conclusions} \label{sec:conclusions}

In this work, we utilised the combined strength of NIRCam imaging and follow-up NIRSpec spectroscopy to analyse the properties of four EoR galaxies observed in the \emph{JWST} Early Release Observations of the  SMACS 0723 field. Our analysis focussed on both the photometric and spectroscopic properties of these sources, aiming to draw on the synergy between imaging and spectroscopy. As the spectroscopic redshifts of these systems were known, we were able to undertake a preliminary investigation into the actual prospects of identifying such sources from SED fitting and colour--colour selections applied to NIRCam photometry. Through the spectroscopic analysis of the observed emission lines, we were able to draw comparisons against the [\ion{O}{III}] $\lambda 5007$ equivalent widths inferred from the measured F356W$-$F444W colours. Furthermore, we showed how a combination of photometry and spectroscopy could help distinguish between nebular emission driven by star formation and AGN activity. Our main results are as follows.

Through our analysis of the morphological properties of these galaxies, we find that these systems are clearly resolved with NIRCam. Barring any distortions from gravitational lensing, these galaxies have distinct, irregular shapes with bright clumps. Thus galaxies in the epoch of reionisation as seen through NIRCam have neither point-source, nor uniform morphologies, implying that the standard flux calibration procedures for reducing NIRSpec spectra will not be optimal for even the most distant galaxies.

Given that the spectroscopic redshifts of these galaxies are now known, we aim to establish the accuracy with which the photometric redshifts for these objects could be derived, using only the photometry in the six NIRCam bands available. We find that the galaxies at $z_\mathrm{spec} = 7.659, 7.664, 8.495$ are `misidentified' at $z\sim10$, with the SED fitting code {\tt BAGPIPES} incorrectly attributing the very red F356W$-$F444W colour ($\geq 0.6$~mag) to a strong Balmer break spanning these two filters, rather than being due to high EW [\ion{O}{III}] $\lambda 5007$ emission in the F444W filter at $z=7$--$9$.

Indeed, we showed that with the addition of imaging with F115W, one should be able to more effectively distinguish between these two redshift regimes, through the presence or absence of a Lyman break in the F115W filter. 

Furthermore, we also investigate how effective colour--colour selections are at identifying these spectroscopically-confirmed high-redshift sources. We find that these galaxies do not exhibit distinct colours in the (F090W$-$F150W) vs.~(F150W$-$F200W) colour plane, which is based on a Ly$\alpha$ break and continuum slope selection, in the former and latter filter pairs, respectively. In contrast, these sources all displayed extreme (F356W$-$F444W) colours, indicative of high EW [\ion{O}{III}]~$\lambda 5007$ and H$\beta$ emission in either the F444W filter (for the three $z > 7$ sources) or the F356W filter (for the single $z=6.4$ source). Indeed, despite being selected for spectroscopic follow-up from only 5~arcmin${^2}$ of NIRCam imaging, these sources all exhibit (F356W$-$F444W) colours located at the extreme end of theoretical predictions, both from the \citet{Yung2022} semi-analytic model (which spans 750~arcmin${^2}$) as well as the {\footnotesize FLARES} \citep[][]{Lovell2021, Vijayan2021, Roper2022, Vijayan2022, Wilkins2022a, Wilkins2022b} simulations (40 zoom-in simulations drawn from a parent volume of 3.2$^3$~cGpc$^3$). Hence it is likely that galaxies in the epoch of reionisation have, on average, even stronger emission lines than these models predict, even when modelling the most extreme environments.

On the spectroscopic side we analysed the $R\sim1000$ spectra taken by the NIRSpec/G395M grating, which spans the spectral range $2.9 < \lambda\ ($\textmu m$) < 5.3$. At the redshifts of the galaxies studied in this work, this observed wavelength range corresponds to the rest-frame optical. Thus we are able to apply standard spectroscopic emission line diagnostics, but now applied to galaxies deep within the epoch of reionisation. Given current uncertainties regarding the accuracy of the NIRSpec flux calibration as a function of wavelength, we focused on studying line ratios of lines at comparable wavelength.

We find that these EoR galaxies deviated (by $\approx$ 0.5~dex) from the local correlation between the [\ion{O}{III}] $\lambda 5007$/H$\beta$ and [\ion{Ne}{III}] $\lambda 3869$/[\ion{O}{II}] line ratios. We attributed this deviation to an excess in the [\ion{Ne}{III}] $\lambda 3869$/[O II] line ratio, which is consistent with the predictions from the {\footnotesize FLARES} simulations, where this offset is likely driven by the lower metallicity in galaxies at $z =$ 6–9. Furthermore, from a measurement of the line flux and continuum level, we find that these galaxies exhibit large (400--1000~\AA) [\ion{O}{III}] $\lambda 5007$ rest-frame equivalent widths, which are consistent with the large equivalent widths inferred from the (very red) measured (F356W$-$F444W) colours. We find that the measured [\ion{O}{III}] $\lambda 5007$ and H$\beta$ EWs are consistent with those seen in extreme, intensely star-forming dwarf galaxies in the local Universe. However, in line with the {\footnotesize FLARES} predictions, such intense star formation and extreme nebular conditions are likely the norm, rather than the exception, in the epoch of reionisation.

Finally we discuss the prospects for distinguishing between star-forming galaxies and AGN in the epoch of reionisation using spectroscopy. Applying the [\ion{O}{III}] $\lambda 5007$/H$\beta$, [\ion{O}{II}]/H$\gamma$ BPT diagram, the emission lines for which are all still accessible by NIRSpec at $z > 7$, we are unable to distinguish between nebular emission driven by star formation and AGN in our four EoR galaxies. However, upon using the colours from NIRCam photometry, and assuming the \citet{Nakajima2022} models for star-forming galaxies and AGN, we find that star-forming galaxies and AGN exhibit distinct colours in the (F150W$-$F200W) vs. (F200W$-$F277W) plane. This separation in colour is attributable to the (assumed) power law continuum for AGN, which results in red colours in the aforementioned filter pairs at $7 < z < 9$. 

Thus it is through the combination of photometry and spectroscopy, by drawing on their synergy, that we will be able to obtain the most clear and comprehensive account of the properties of galaxies in the epoch of reionisation during the \emph{JWST} era. 

\section*{Acknowledgements}

We thank the referee for their useful comments which helped to improve this article. JT thanks Andrew Bunker and Roberto Maiolino for useful discussions that helped guide the direction of this article. JT also thanks Tom Broadhurst for confirming that galaxies 06355 and 10612 are two distinct (i.e.\@ not multiply-imaged) sources. 

We are grateful to acknowledge Anthony Holloway, Sotirios Sanidas and Phil Perry for critical and timely help with computer infrastructure that made this work possible.
We acknowledge support from the ERC Advanced Investigator Grant EPOCHS (788113), as well as a studentship from STFC.  This work is based on observations made with the NASA/ESA \textit{Hubble Space Telescope} (HST) and NASA/ESA/CSA \textit{James Webb Space Telescope} (JWST) obtained from the \texttt{Mikulski Archive for Space Telescopes} (\texttt{MAST}) at the \textit{Space Telescope Science Institute} (STScI), which is operated by the Association of Universities for Research in Astronomy, Inc., under NASA contract NAS 5-03127 for JWST, and NAS 5–26555 for HST. These observations are associated with program 14096 for HST, and 2736 for JWST. LF acknowledges financial support from Coordenação de Aperfeiçoamento de Pessoal de Nível Superior - Brazil (CAPES) in the form of a PhD studentship. CCL acknowledges support from the Royal Society under grant RGF/EA/181016. The Cosmic Dawn Center (DAWN) is funded by the Danish National Research Foundation under grant No. 140.

This research made use of Astropy,\footnote{http://www.astropy.org} a community-developed core Python package for Astronomy \citep{astropy2013, astropy2018}. 

\section*{Data Availability}

The NIRCam and NIRSpec data used in this analysis is publicly available from the Mikulski Archive for Space Telescopes (MAST) website: https://mast.stsci.edu/portal/Mashup/Clients/Mast/Portal.html. Any remaining data underlying the analysis in this article will be shared on reasonable request to the first author.



\bibliographystyle{mnras}
\bibliography{main.bib} 




\appendix

\section{Colour selection}

\begin{figure*}
\centering
\includegraphics[width=\linewidth]{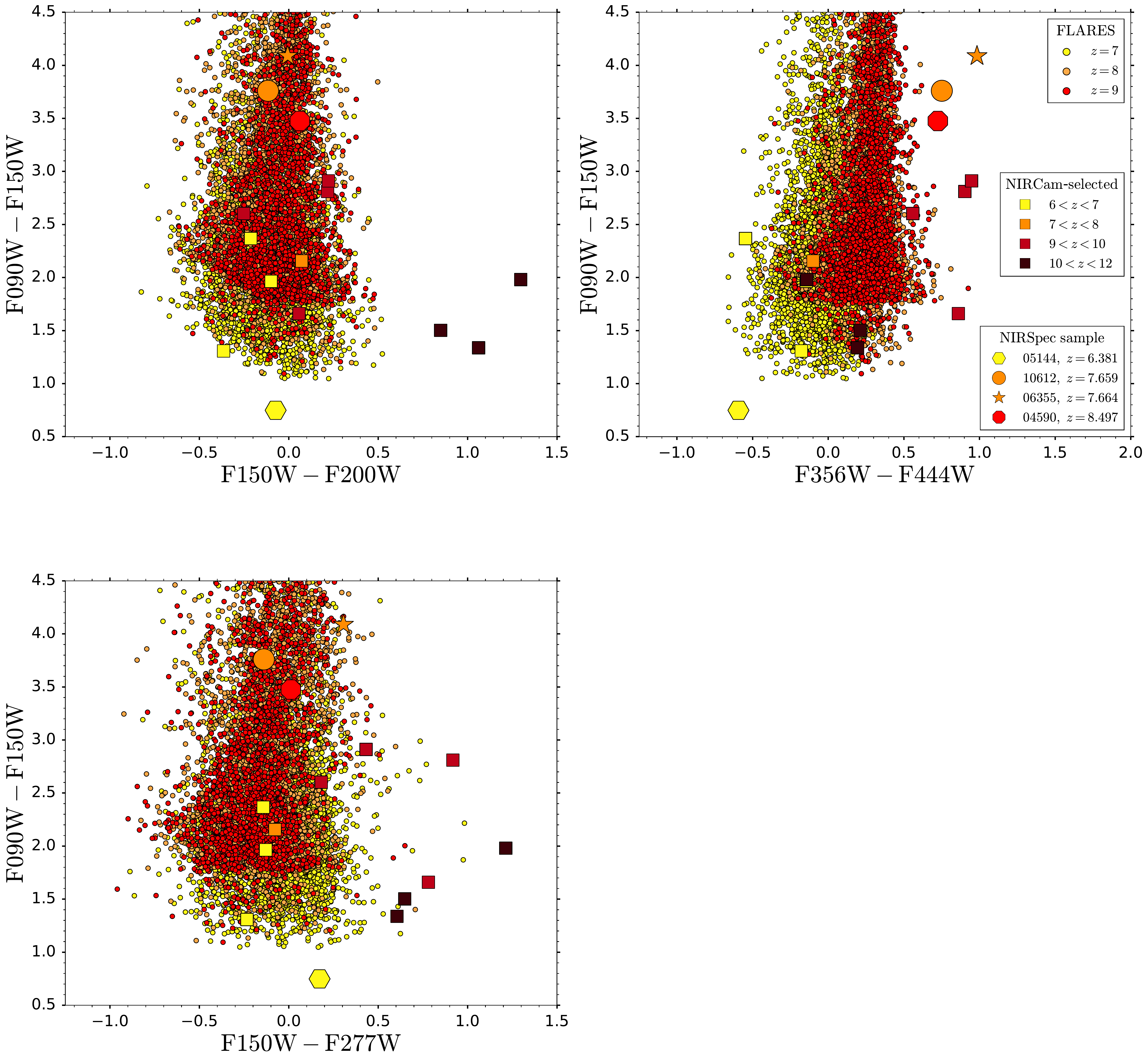}
\caption{Similar to Fig.\@~\ref{fig:colours_sam}, but now showing the NIRCam colours predicted for $z=7,\ 8,\ 9$ (yellow, orange, and red, respectively) galaxies by the FLARES simulation. Although the FLARES simulation includes stronger emission lines, and thus redder F356W$-$F444W colours than the \citet{Yung2022} semi-analytic model, we still find that the four epoch of reionisation galaxies lie at the extreme ends of the simulation predictions.}
\label{fig:colours_flares}
\end{figure*}

In the main body of the paper we showed the colours predicted by the \citet{Yung2022} semi-analytic model, in comparison to the colours of the four EoR galaxies with spectroscopic redshifts. In Fig.\@~\ref{fig:colours_flares} we instead show the predictions from individual snapshots of the {\footnotesize FLARES} simulations \citep[][]{Lovell2021, Vijayan2021, Vijayan2022, Roper2022, Wilkins2022a, Wilkins2022b}. We do note that interpolating between snapshots can lead to more extreme colour evolution, as emission lines pass in and out of photometric bands \citep[see][]{Wilkins2022b}, which is an effect accounted for in the \citet{Yung2022} lightcones. We find once again that the F356W$-$F444W colours (sensitive to the EW of [\ion{O}{III}] $\lambda 5007$ and H$\beta$) of our four EoR galaxies (selected from only 5~arcmin$^{2}$ of NIRCam imaging), lie within the extreme ends of the {\footnotesize FLARES} simulation predictions (where the 40 zoom-in simulations have been selected from a parent volume of 3.2~cGpc$^3$, thus capturing rare overdense regions and extreme objects in colour space). Thus our results indicate that galaxies in the epoch of reionisation may, on average, have stronger emission lines than the {\footnotesize FLARES} simulations predict.

\section{Photometric properties}\label{app:photometry}

In Section~\ref{sec:photometry} we used the SED-fitting code {\tt BAGPIPES} to investigate the accuracy with which photometric redshfits can be derived using NIRCam photometry alone. In this Appendix, we repeat this analysis using {\tt Le Phare} and {\tt EAZY}. We follow the fitting methodology outlined in \citet{Adams2023} and \citet{Austin2023}. 

\begin{figure*}
\centering
\includegraphics[width=0.55\linewidth]{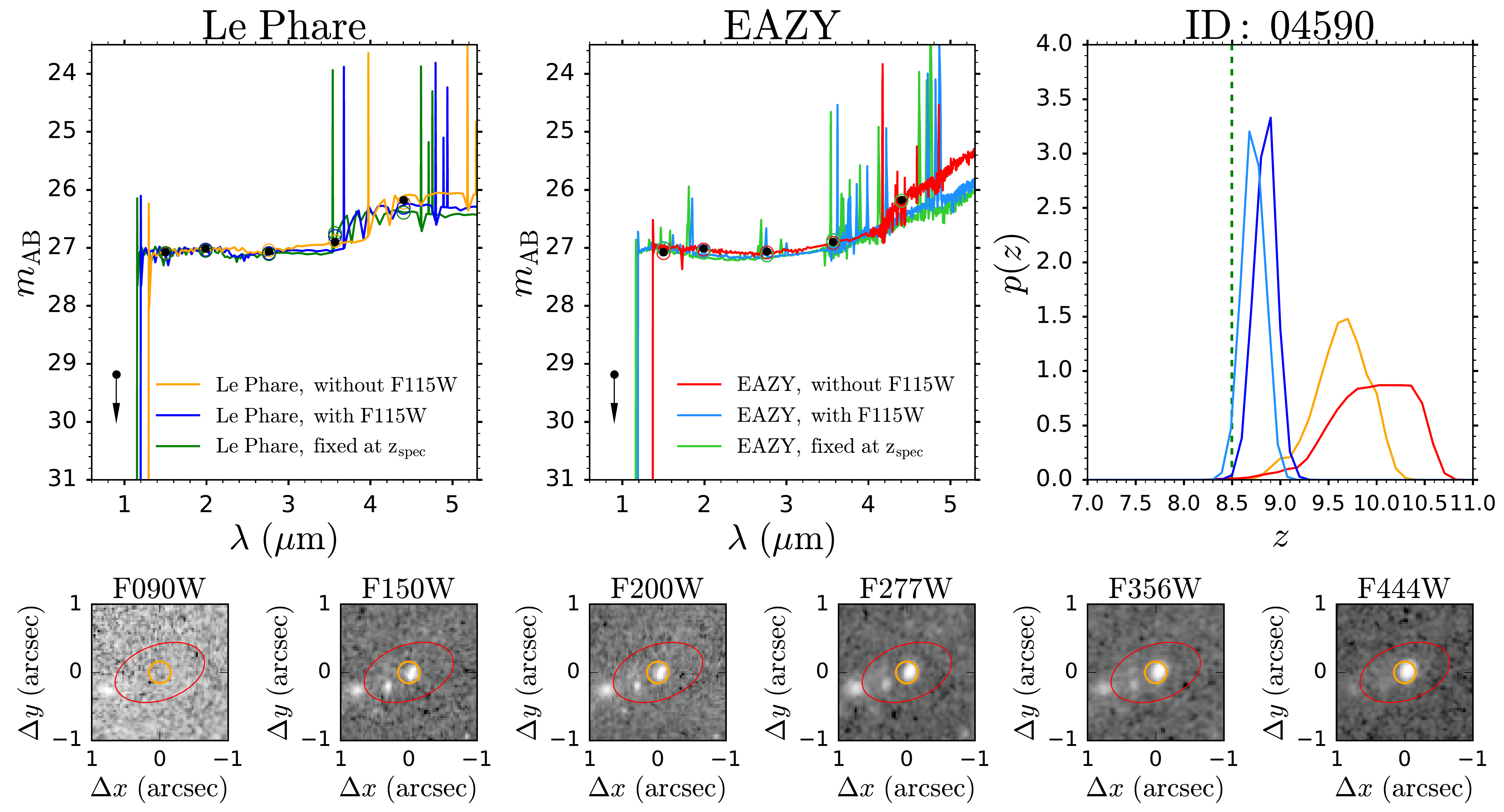} \\[5ex]
\includegraphics[width=0.55\linewidth]{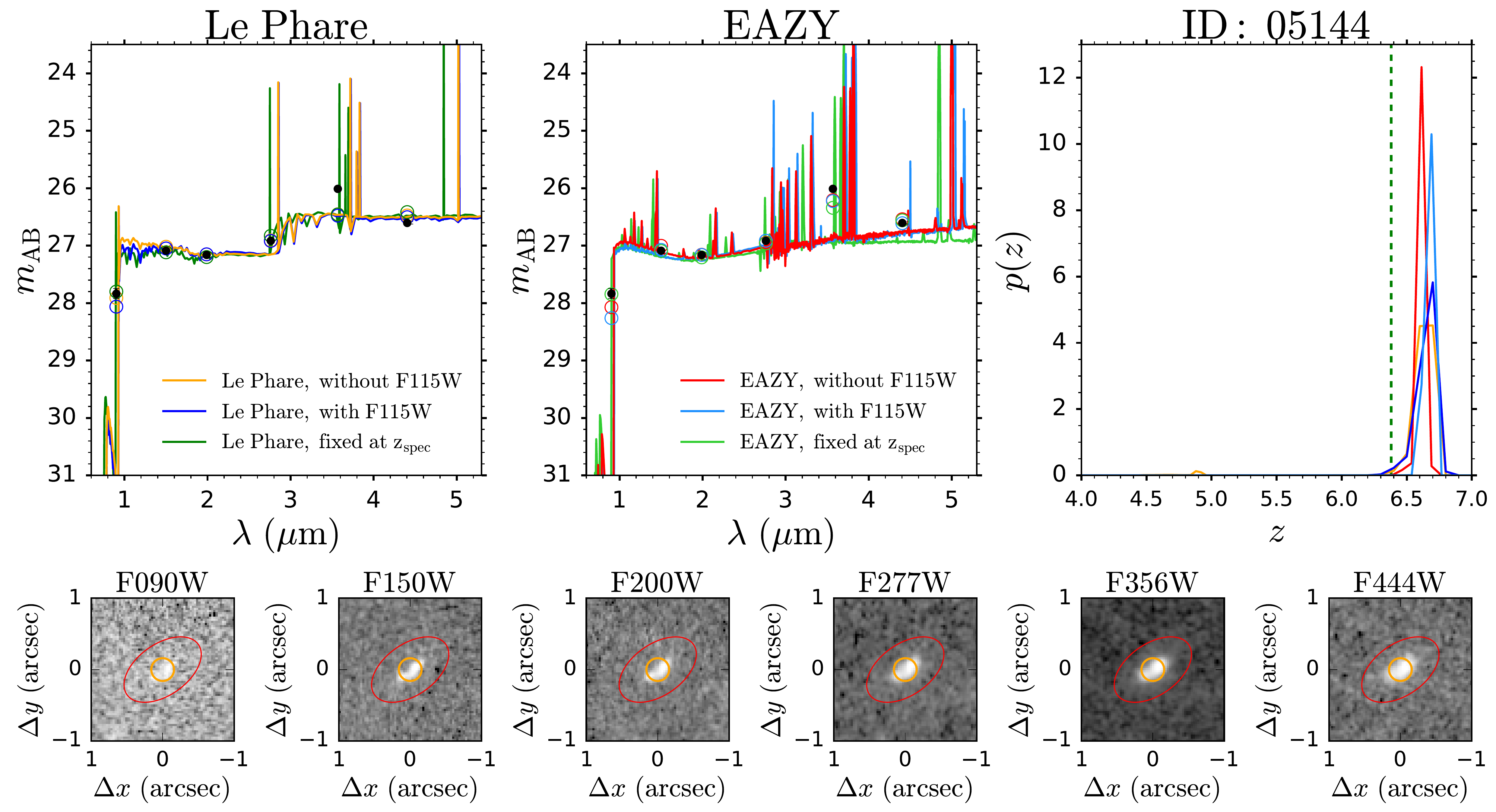} \\[5ex]
\includegraphics[width=0.55\linewidth]{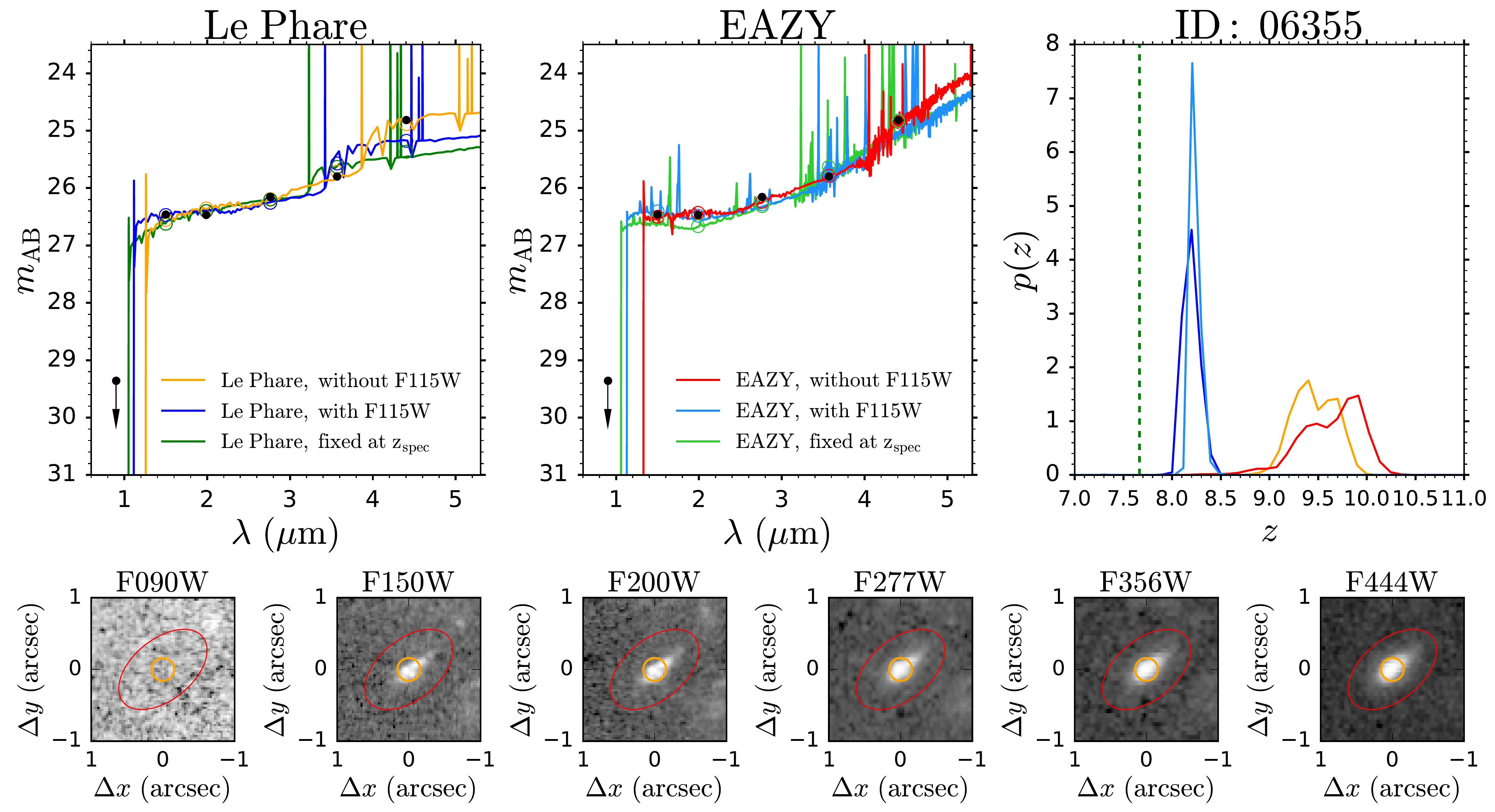} \\[5ex]
\includegraphics[width=0.55\linewidth]{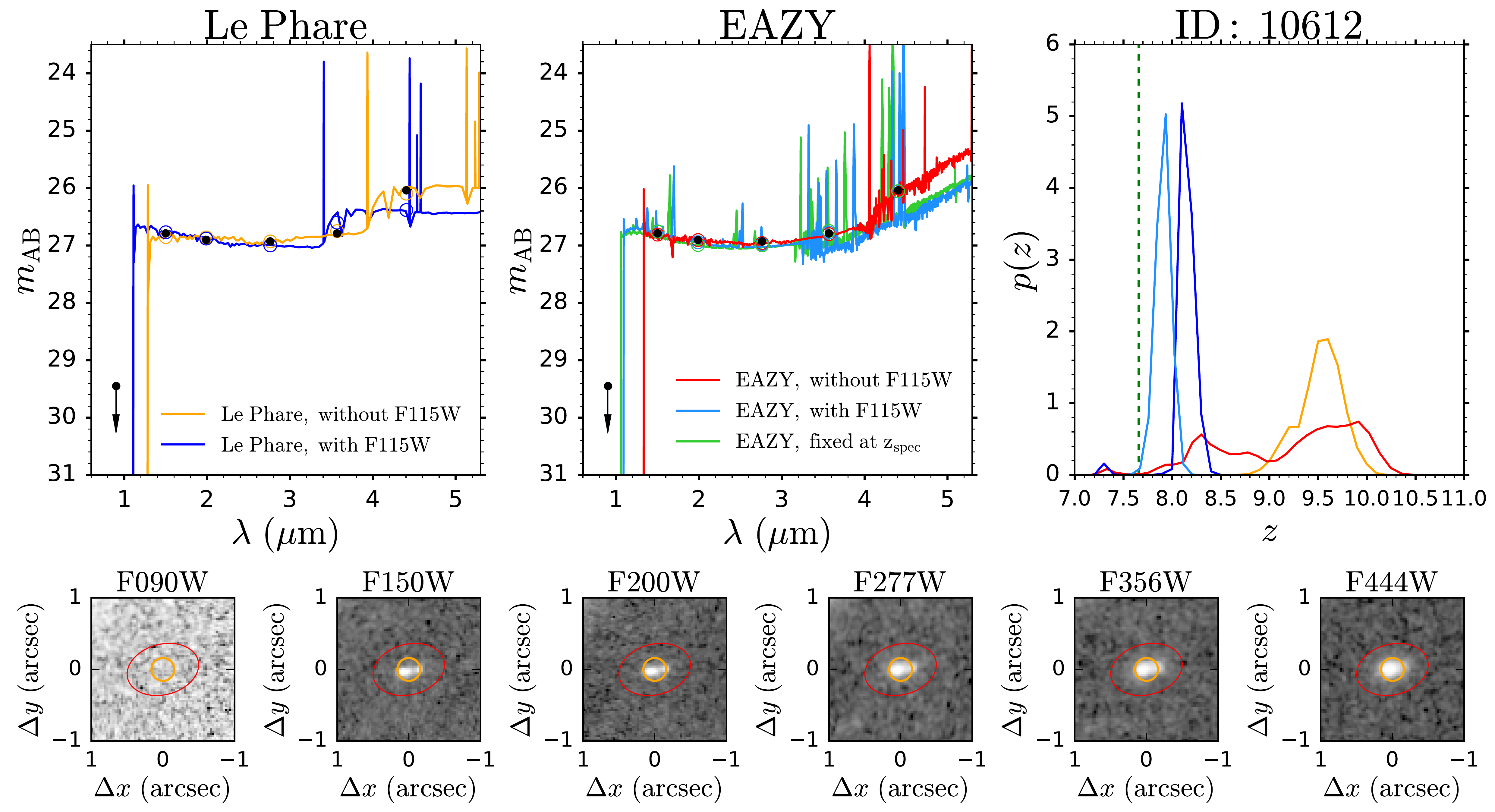}
\caption{SED fits with Le Phare (top-left panel) and EAZY (top-middle panel) to the available NIRCam photometry (F090W, F150W, F200W, F277W, F356W, F444W, shown in black) for the four EoR galaxies. Similar to Fig.\@~\ref{fig:bagpipes_fits}, we show the best fits obtained using the available photometry (i.e.\@ without F115W, orange/red) including F115W (dark/light blue) and keeping the redshift fixed at the known spectroscopic redshift (dark/light green). Top-right panel: The posterior redshift probability distributions. Bottom panels: Cutouts for the six NIRCam bands, together with the 0.32 arcsec diameter extraction aperture (used for SED fitting and colour analysis, orange) and Kron elliptical aperture (used to derive the total stellar masses and SFRs, red).}
\label{fig:lp_eazy_cutouts}
\end{figure*}

The observed 0.32 arcsec aperture-corrected photometry, together with the best-fit {\tt Le Phare} (left panels) and {\tt EAZY} (middle) SEDs, as well as the resulting posterior redshift probability distributions (right) are shown in Fig.\@~\ref{fig:lp_eazy_cutouts}. We also show cutouts for the six NIRCam bands, together with our 0.32 arcsec diameter (orange) and Kron elliptical extraction apertures (red).

We note that the flux extracted from the Kron elliptical aperture for galaxy 04590 is somewhat contaminated by the light from two neighbouring sources on the sky. The stellar mass and SFR determined for this galaxy using the 0.32 arcsec aperture-corrected fluxes are instead $\log (M_*\mathrm/{M}_\odot$) = 7.44 and SFR = 0.28~$\mathrm{M_\odot}$~$\mathrm{yr^{-1}}$.

As with our {\tt BAGPIPES} analysis, without a F115W measurement (orange and red curves), we find that the flux excess in the F444W band is incorrectly attributed to a $z\sim10$ Balmer break, rather than $z\sim8$ [\ion{O}{III}] + H$\beta$ line emission. By including mock F115W photometry, both the {\tt Le Phare} and {\tt EAZY} photometric redshifts (dark and light blue) better agree with the known spectroscopic redshifts. However, there are usually still some discrepancies, with e.g.\@ the posterior redshift distributions having zero probability at the known spectroscopic redshift (ID 06355), the {\tt Le Phare} fits fixed at the known spectroscopic redshift (shown in green) not having sufficient emission line strength to match the data (IDs 05144, 06355), or the {\tt Le Phare} fit fixed at the spectroscopic redshift failing entirely (10612). Thus F115W imaging, together with templates which incorporate high equivalent width emission lines, are likely both essential in order to derive accurate photometric redshifts for EoR galaxies with NIRCam data \citep[see also][]{Adams2023}.

\bsp	
\label{lastpage}
\end{document}